\documentclass{article}
\usepackage{amssymb}
\usepackage{amsbsy}

\usepackage{comment}

\usepackage{enumerate}
\usepackage{graphicx}
\newcommand{\be}{\begin{equation}}
\newcommand{\ee}{\end{equation}}

\def\bea{\begin{eqnarray}}
\def\eea{\end{eqnarray}}
\def\bean{\begin{eqnarray*}}
\def\eean{\end{eqnarray*}}

\newcommand{\barr}{\begin{array}}
\newcommand{\earr}{\end{array}}

\newcommand{\bed}{\begin{displaymath}}
\newcommand{\eed}{\end{displaymath}}
\newcommand{\bal}{\begin{array}{ll}}
\newcommand{\eal}{\end{array}}

\def\bvec#1{\raise1.5ex\hbox{$\rightarrow$}\mkern-16.5mu #1}

\def\ket#1{\vert\,#1\rangle}
\def\m#1{\mathcal#1}

\newcommand{\bs}{\boldsymbol}

\begin{document}

\title{\hfill ~\\[-30mm]
       \hfill\mbox{\small }\\[30mm]
       \textbf{Majorana Physics Through the \\ Cabibbo Haze}
       %Higgs in the $\boldsymbol{\m Z_7 \rtimes \m Z_3}$ Family}
       } 
\date{}
\author{\\        Jennifer Kile,\footnote{E-mail: {\tt jenkile@phys.ufl.edu}}~~M. Jay P\'erez,\footnote{E-mail: {\tt mjperez@phys.ufl.edu}}~~Pierre Ramond,\footnote{E-mail: {\tt ramond@phys.ufl.edu}}~~Jue Zhang,\footnote{E-mail: {\tt juezhang@phys.ufl.edu}}\\ \\
  \emph{\small{}Institute for Fundamental Theory, Department of Physics,}\\
  \emph{\small University of Florida, Gainesville, FL 32611, USA}}

\maketitle

\begin{abstract}
\noindent {\em} We present a model in which the Supersymmetric Standard Model is augmented by the family symmetry $\bs{\m Z_7 \rtimes \m Z_3}$. Motivated by $SO(10)$, where the charge two-thirds and neutral Dirac Yukawa matrices are related, we propose, using family symmetry, a special form for the seesaw Majorana matrix; it contains a squared correlated hierarchy, allowing it to mitigate the severe hierarchy of the quark sector. It is reproduced naturally by the invariant operators of $\bs{\m Z_7 \rtimes \m Z_3}$, with the hierarchy carried by familon fields. In addition to relating the hierarchy of the $\Delta I_{\rm w}=1/2$ to the $\Delta I_{\rm w}=0$ sector, it contains a Gatto-Sartori-Tonin like relation, predicts a normal hierarchy for Tri-bimaximal and Golden Ratio mixings, and gives specific values for the light neutrino masses. 

\end{abstract} 

\thispagestyle{empty}
\vfill
\newpage
\setcounter{page}{1}

%%%%%%%%%%%%%
\section{Introduction}
LHC's discovery of the Standard Model's Higgs closes a chapter in its original formulation, while both dark matter and massive neutrinos\cite{whitepaper} indicate an incomplete description of Nature and open two new chapters, ``Physics Beyond the Standard Model", and ``Physics Well Beyond the Standard Model". 

The search for BSM physics proceeds  by direct searches for new massive particles;  for WBSM physics it proceeds through the detection of rare processes, including proton decay, and precision measurements of neutrino masses and mixings.

An important legacy of the Standard Model is the quark-lepton unification \cite{GUT} at large energies (``GUT scale"), a pattern that is consistent both with the gauge quantum numbers of the three matter families, and the near convergence of its three couplings at large energy.  This implies also several  relations between quark and (at least) charged lepton masses, with mixed success. The origin and values of fermion masses remains a mystery well worth investigating.\footnote{{\em ``You want more than a Nobel Prize? You want to become a king? Figure that out!''} (R.P. Feynman, 1978)}

Mixing the left-handed neutrinos with Majorana neutrinos with masses at GUT scales provides a natural explanation for small neutrino masses\cite{seesaw,lseesaw}, but the principle(s) which determines the right-handed Majorana neutrino masses and mixing patterns is unknown physics of the WBSM variety.

Moreover, physics at  GUT scale runs into two conceptual difficulties:  One, the values of the Higgs and top quark masses suggest a breakdown of the Standard Model's potential at energies orders of magnitudes lower than the GUT scale; second, quark-lepton unification does not naively apply to all masses and mixings, since the striking hierarchy of the quarks is not seen in neutrinos.  

The first conceptual hurdle can be finessed in the Supersymmetric Standard Model (SSM) where the  potential is no longer unstable, and has the added benefit of the remarkable convergence of the three gauge coupling constants\cite{gaugeunif}: the implementation of grand-unified theories requires supersymmetry.

The second hurdle is not so clearly overcome. We investigate in this paper a model in which the SSM is supplemented with the family symmetry $\bs{\m Z_7 \rtimes \m Z_3}$, the smallest  discrete non-Abelian subgroup of $SU(3)$.  Its effect is to restrict  flavor patterns and, together with GUT-like assumptions, to relate the light neutrino mixing and mass patterns with the seesaw GUT-scale Majorana matrix. 
% The proper choice of  family symmetry can also shed some light on hierarchies generated  from the vacuum. 
%We assume the seesaw mechanism, where the neutrino masses depend upon both high-scale Majorana mass terms and weak-scale Dirac mass terms. 

Motivated by GUT ideas, we take the neutral Dirac mass matrix to show the same hierarchy as the up quarks. Since neutrinos do not show this extreme hierarchy, the Majorana matrix must itself be hierarchical so as to cancel that of the Dirac matrix. 

We find one remarkable Majorana matrix compatible with $\mu$-$\tau$ symmetry. It is most elegantly generated by a single dimension-six operator with three familon fields, although it may also be obtained from special linear combinations of dimension-five operators.  

In the original version of this paper, we mistakenly identified a particular linear combination of dimension-five operators and motivated it by an underlying theory. As this linear combination does not produce the Majorana matrix discussed in the text, we have deleted the relevant section in this revised version.

The familon vacuum values can explain not only the hierarchy of this Majorana matrix, but also relate it to that of the quarks, at the expense of introducing unknown familon physics. In some sense this is a ``bottom-up'' approach, where we hope that a particularly interesting operator can shed light on the physics of the familon sector. 

This special matrix predicts the normal hierarchy, with Tri-bimaximal mixing for the seesaw mixing angles, and specific values for the neutrino masses. However, comparison with neutrino data requires a knowledge of the matrix which diagonalizes the charged lepton Yukawa matrix. By GUT ideas it is controlled by Cabibbo effects, forming a ``Cabibbo Haze'' through which the neutrino parameters must be inferred. In particular, the value of $\theta_{13}$ must be generated by the charged lepton sector.

%%%%%%%%%%%%%%%%%%%%%%%%%%%
\section{The Family Group $\bs{\m Z_7 \rtimes \m Z_3}$}

The choice of family symmetry is dictated by the extreme quark and charged lepton hierarchies,

\be\label{hierarchy}
\begin{pmatrix}{0&0&0\cr  0&0&0\cr 0& 0&1}\end{pmatrix}.\ee
This pattern  with two zeros on the diagonal suggests an $SU(3)$ rather than $SO(3)$ or $SU(2)$ family symmetry. This follows by considering the bilinear products of their respective triplet representations. 

For $SO(3)$, the symmetric product of two triplets breaks into a symmetric traceless matrix ($\bf{5}_s$) and the trace ($\bf{1}$). The zeros of Eq.(\ref{hierarchy}) must then be produced by an exact conspiracy between Yukawa couplings. Such a trace decomposition does not occur in $SU(3)$\cite{SU3}. 
 
Finite subgroups of $SU(2)$  which have both doublet (for quarks and charged leptons), and triplet representations (for neutrinos),  do not treat leptons and quarks on the same footing, and are anathema to the spirit of quark-lepton unification. 

We therefore focus on $\bs{\m Z_7\rtimes\m Z_3}$, the smallest non-Abelian subgroup of $SU(3)$. 
%The reader will find a summary of its important group-theoretic properties in Appendix A. 
As can be seen in Appendix A, this twenty-one element group has, besides a real singlet representation, a complex  triplet, a complex one-dimensional representation, and their conjugates. Its distinguishing feature is that Yukawa couplings between two triplets will either be
completely diagonal or completely off-diagonal. 

We consider as our background a unified framework in which the quarks and leptons are related, such as $SU(5)$ or $SO(10)$. To avoid hierarchy problems and naturally single out particular couplings, we will also assume our model to be supersymmetric. {\it All} $SU(5)$ matter superfields,  $\bf \overline  5$($\psi$), $\bf{10}$($\chi$), and $\bf 1$($N$), transform as $\bs{\m Z_7\rtimes\m Z_3}$ triplets, an assignment that easily generalizes to the $\bf 16$ of $SO(10)$. 

According to $\bs{\m Z_7 \rtimes \m Z_3}$, the quadratic combinations of two matter fields transform either as family triplets or anti-triplets, requiring Higgs anti-triplets or triplets, respectively.  This differs from an earlier model\cite{z7z3singlet} in which the Higgs fields  ${\m H}^{}_{u,d}$ were ${\bs{\m Z_7\rtimes\m Z_3}}$ singlets. 

If  the Higgs fields were family triplets, the matter field bilinears would combine into antitriplets, $(\chi\chi)^{}_{\bf \overline  3},(\psi\chi)^{}_{\bf \overline  3}$, and $ (\psi N)^{}_{\bf \overline  3}$,  with only off-diagonal elements, and thus with unwanted tree-level mass degeneracies. 

The assignments which reproduce the hierarchy Eq.(\ref{hierarchy}) require both ${\m H}^{}_{u,d}$ to be antitriplets ($\bf \bar 3$), and the matter field bilinears to be triplets. The tree-level superpotential

\be \label{eq:potential}
W = y^{}_{10}(\chi\chi)^{}_{\bf 3}{\m H}^{}_{u} +  y^{}_{\bar{5}} (\psi\chi)^{}_{\bf 3}{\m H}^{}_{d} + y^{}_{1} (\psi N)^{}_{\bf 3}{\m H}^{}_{u} + \ldots,
\ee
reproduces the extreme quark hierarchy with the simple (approximate) Higgs vacuum,

\be \label{huvev}
\langle {\m H}^{}_{u,d} \rangle^{}_0 ~=~\begin{pmatrix}{v^{}_{u,d}\cr 0\cr 0}\end{pmatrix}.\ee
As pointed out in \cite{z7z3singlet}, such a vacuum alignment, along a single direction in family space, can be naturally accommodated in $\bs{\m Z_7 \rtimes \m Z_3}$.

Each Standard Model Higgs field now has two family partners. We do not take up the question of their role in this publication, and assume them to be heavier copies of the normal Higgs.\footnote{In an earlier publication\cite{s3}, we investigated their role as messengers of supersymmetry breaking in a toy model with only one Higgs family partner, assuming an $\bs{\m S}_3$ family symmetry group.}  Instead, we analyze the question of whether $\bs{\m Z_7\rtimes\m Z_3}$ can provide a natural framework for reconciling the extreme hierarchy of the charge (2/3) sector with the mild hierarchy of the neutrinos, and also shed light on the seesaw Majorana neutrino mass matrix. 

%%%%%%%%%%%%%%%%%%%%%%%%%%%%%
\section{Majorana Physics}

Since the right-handed neutrinos ($N$) have no electroweak quantum numbers, a question that comes to mind is, ``What sets the physics of the Majorana neutrinos?'' A natural answer is family symmetry. From this point of view, the Majorana mass matrix $\m M$ is a window to family symmetry, and it is a task of theory to ``extract" its structure from the measured masses and mixing patterns of the three light neutrinos. 

The seesaw mechanism, with a GUT-scale $\Delta I_{\rm w}=0$ Majorana matrix ${\m M}$, and the neutral $\Delta I_{\rm w}=1/2$ Dirac Yukawa matrix $Y^{(0)}$,   generates small neutrino masses, as well as mixing of the neutrino flavors through the relation,

\be \label{eq:seesaw}
M^{}_\nu =  Y_{}^{(0)} \m M_{}^{-1} Y_{}^{(0) T}.
\ee
Masses and mixings are derived through the diagonalization of $M_\nu$, 

\begin{equation}
M^{}_\nu = \m U^{}_{\rm \,seesaw}~ D^{}_\nu~ \m U^T_{\rm \,seesaw},
\end{equation}
where  $D_\nu={\rm diag}(m_1,m_2,m_3)$ is the diagonal neutrino mass matrix, and 
 $\m U_{\rm \,seesaw}$ is the seesaw neutrino mixing matrix. Although in general two Majorana phases can appear in $D_\nu$, we choose to put them aside in this paper and assume all $m$'s to be real. Absorbing unphysical phases, 
 
 \be
{\m U}^{}_{\rm \,seesaw}~ = ~\begin{pmatrix}{1&0&0\cr 0&c^{}_{23}&-s^{}_{23}\cr 0&s^{}_{23}&c^{}_{23}}\end{pmatrix}
\begin{pmatrix}{c^{}_{13}&0&-e^{-i\delta}_{}s^{}_{13}\cr 0&1&0\cr e^{i\delta}_{}s^{}_{13}&0&c^{}_{13}}\end{pmatrix}
\begin{pmatrix}{c^{}_{12}&-s^{}_{12}&0\cr s^{}_{12}&c^{}_{12}&0\cr 0&0&1}\end{pmatrix},
\ee
contains one Dirac CP-violating phase $\delta$ and three seesaw mixing angles, denoted by $\eta_{12}$, $\eta_{23}$ and $\eta_{13}$ 
 ($s_{ij}=\sin\eta_{ij}$,   $c_{ij}=\cos\eta_{ij}$), to distinguish them from their measured counterparts $\theta_{ij}$ in the observable MNSP matrix,

\be
\m U^{}_{MNSP}~=~\m U^\dagger_{-1}\,\m U^{}_{\rm seesaw},\ee 
with $\m U_{-1}$ determined by the charged lepton Yukawa matrix $Y^{(-1)}$. In GUT theories, where Yukawa Dirac matrices of quarks and charged leptons are  related,  we expect the CKM parameters to appear in neutrino mixings as well, forming a ``Cabibbo Haze"\cite{haze} between the data and the Majorana matrix.   

With the recent Daya Bay\cite{dayabay}, RENO\cite{reno} and Double Chooz\cite{dchooz} measurements of the reactor angle, all three mixing angles have now been measured, and a global fit\cite{fit}, with a one $\sigma$ error, gives 

\be \label{eq:angles}
\theta^{}_{12} = {33.6^\circ}^{+1.2^\circ}_{-1.0^\circ}, \qquad \theta^{}_{23} = {38.4^\circ}^{+1.4^\circ}_{-1.2^\circ}, \qquad \theta^{}_{13} = {8.9^\circ}^{+0.5^\circ}_{-0.4^\circ}.
\ee 
Although the light neutrino masses remain unknown, their mass differences are determined by neutrino oscillation experiments,

\be 
\Delta m_{21}^2 = {7.54}^{+0.26}_{-0.22} \times 10^{-5}~eV^2, \qquad |\Delta m_{31}^2| = 2.43^{+0.06}_{-0.10} \times 10^{-3}~eV^2.
\ee
The sign of $\Delta m_{31}^2$, yet to be measured, distinguishes the normal ($+$) and inverted ($-$) hierarchies. This, together with the cosmological bound from Planck\cite{Planck},

$$
\sum m_{i} \leq 0.23~eV ,$$
show that the neutrinos do not display the same hierarchy of the quarks. 

This discrepancy in hierarchies is quite puzzling from a grand unified point of view, where quark and lepton matrices are related. A resolution within the framework of the seesaw mechanism is that the Majorana matrix contains a squared correlated hierarchy. 

Since in $SO(10)$, the neutral Dirac Yukawa matrix is naturally related to that of the up-quarks,  we assume they both display the same hierarchy, 

\begin{eqnarray}
Y^{(2/3)}_{}\sim Y^{(0)}_{}\sim \begin{pmatrix}{\lambda^8_{} & 0 & 0\cr 0&\lambda^4_{}& 0\cr 0& 0& 1}\end{pmatrix},
\end{eqnarray}
compatible with the up quark masses, parametrized in terms of the Cabibbo angle at the GUT scale\cite{ross&serna}, $\lambda = \sin \theta_c = 0.227$. To undo the hierarchy in $Y^{(0)}$, we must then have,

\be
{\m M}~\sim~\begin{pmatrix}{a^{}_{11}\lambda^{16}_{}&a^{}_{12}\lambda^{12}_{}&a^{}_{13}\lambda^8_{}\cr a^{}_{12}\lambda^{12}_{}&a^{}_{22}\lambda^8_{}&a^{}_{23}\lambda^4_{}\cr a^{}_{13}\lambda^8_{}&a^{}_{23}\lambda^4_{}& a^{}_{33}}\end{pmatrix},\ee 
where all $a_{ij}$ are of order one.

The Majorana mass matrix $\m M$ plays dual roles. On the one hand, it is very hierarchical, with the {\em same} factor $\lambda^4$ that appears in the up-quark Yukawa matrix. This allows it to undo the hierarchy of the Dirac Yukawa matrix, resulting in a mild spectrum in the neutrino sector. On the other hand, its pre-factors are directly related with the neutrino mass matrix, and encode the measured neutrino masses and mixings. 

A dictionary between the pre-factors and the parameters in the seesaw mixing matrix can then be established in a series of steps.

\vskip .3cm
$\bullet$ \textbf{Step I :}
\vskip .3cm

As remarked by many authors\cite{mutau}, a simple constraint among the pre-factors, ($2-3$ or $\mu-\tau$ symmetry), determines two of the three seesaw mixing angles,

\begin{eqnarray}
a^{}_{12}=a^{}_{13},\qquad a^{}_{22}=a^{}_{33},~~~~\longrightarrow~~~~\eta^{}_{23}=45^\circ,\qquad \eta^{}_{13}~=~0^\circ,
\end{eqnarray}
close to their experimental values, with $\delta$, $\eta_{12}$  and the three masses undetermined.  The pre-factors can then be expressed in terms of neutrino masses and $\eta_{12}$, 

%$${\m M}~=~\begin{pmatrix}{a\lambda^{16}_{}&b\lambda^{12}_{}&b\lambda^8_{}\cr b\lambda^{12}_{}&c\lambda^8_{}&d\lambda^4_{}\cr b\lambda^8_{}&d\lambda^4_{}& c}\end{pmatrix},
%\begin{pmatrix}{ax^4_{}&bx^3_{}&bx^2_{}\cr bx^3_{}&cx^2_{}&dx\cr bx^2_{}&dx& c}\end{pmatrix},$$

\begin{eqnarray}
\label{eq:dictionary}
a^{}_{11}&=& \frac{c^2_{12}}{m^{}_1}+\frac{s^2_{12}}{m^{}_2}, \qquad \qquad \qquad ~
a^{}_{12}~=~ \frac{1}{\sqrt{2}} \Big(\frac{1}{m^{}_1}-\frac{1}{m^{}_2}\Big)c^{}_{12}s^{}_{12},\nonumber \\
a^{}_{22}&=&  \frac{1}{2m^{}_3}+\frac{c^2_{12}}{2m^{}_2}+\frac{s^2_{12}}{2m^{}_1},\qquad 
a^{}_{23}~=  -\frac{1}{2m_3}+\frac{c^2_{12}}{2m^{}_2}+\frac{s^2_{12}}{2m^{}_1}.
\end{eqnarray}
or alternatively, 

$$\frac{1}{m^{}_3} ~=~  a^{}_{22}-a^{}_{23}, \qquad \qquad \frac{1}{2\sqrt{2}}\tan 2\eta^{}_{12} ~=~ \frac{a^{}_{12}}{a^{}_{11}-a^{}_{22}-a^{}_{23}},$$  
$$~~~~\frac{1}{m^{}_1}+\frac{1}{m^{}_2} ~=~ a^{}_{11}+a^{}_{22}+a^{}_{23}, \qquad \Big(\frac{1}{m^{}_1}-\frac{1}{m^{}_2}\Big)^2 ~=~ (a^{}_{11}-a^{}_{22}-a^{}_{23})^2+8a_{12}^2.$$   
  
Neutrino mass hierarchies can also be discussed in terms of pre-factors. They are distinguished by the inequalities,

\bean |a^{}_{22}-a^{}_{23}|&<&|a^{}_{11}+a^{}_{22}+a^{}_{23}|,~~~{\rm normal},\\
  |a^{}_{22}-a^{}_{23}|&>&|a^{}_{11}+a^{}_{22}+a^{}_{23}|,~~~{\rm inverted}.
\eean
There is no one-to-one correspondence between the relative signs of $a_{22}$ and $a_{23}$ and the mass hierarchies. Generically, normal (inverted) hierarchy yield the same (opposite) signs of $a_{22}$ and $a_{23}$, but for the special case,  

$$\frac{c^2_{12}}{2m_2}+\frac{s^2_{12}}{2m_1}=0,\quad \rightarrow~~\tan^2\eta^{}_{12}=-\frac{m^{}_1}{m^{}_2}$$
the normal hierarchy yields $a_{22} = -a_{23}$.

\vskip .3cm
$\bullet$ \textbf{Step II :}
\vskip .3cm

One more relation among the pre-factors yields three popular seesaw mixing matrices  $\m U^{}_{\rm seesaw}$\cite{TBM, GR, flavor}, 

\bean
&&{\rm Tri-bimaximal (TBM)}:   a^{}_{23}~=~ a^{}_{11}- a^{}_{12}- a^{}_{22},\quad~~~\, \tan^2\eta^{}_{12}=\frac{1}{2},\\
&&{\rm Golden~ Ratio (GR)}:~~~~~~   a^{}_{23}~=~ a^{}_{11}-\sqrt{2} a^{}_{12}- a^{}_{22},\quad \tan^2\eta^{}_{12}=\frac{2}{1+\sqrt{5}},\\
&&{\rm Bi-maximal (BM)}:~~~~~~ a^{}_{23}~=~ a^{}_{11}- a^{}_{22}, \qquad \quad~~~~~\, \tan^2\eta^{}_{12}=1.\eean
All three fix the remaining mixing angle, but leave the neutrino masses undetermined. These seesaw mixing matrices have $\eta_{13}=0^\circ$, so that the reactor angle in the MNSP matrix, $\theta^{}_{13}$, along with the necessary corrections needed to bring $\theta_{12}$ and $\theta_{23}$ in agreement with their best fit values in Eq.(\ref{eq:angles}), must be generated by the lepton sector. We provide an example of how this can be accomplished in Appendix C. Also, leptonic CP-violation is determined from that of the quark sector.  

As it stands, the generic mass spectrum of the Majorana matrix displays a severe three-fold hierarchy

$$\Big|\frac{M_1}{M_3}\Big| \sim \m O(\lambda^{16}), \qquad \Big|\frac{M_2}{M_3}\Big| \sim \m O(\lambda^8),$$
which is in tension with bounds from leptogenesis (see \cite{leptogenesis} for a review). In models of leptogenesis where the right-handed neutrino spectrum is very hierarchical, the lightest mass must be larger than $10^8$ GeV\cite{bound}. This would correspond to a spectrum for $\m M$ of $\sim (10^{19}, 10^{14},10^8)$ GeV, pushing the largest mass past the Planck scale. 

There exists a special case for which this hierarchy is mitigated, as can be seen by studying the  eigenvalue ratios of $\m M$; from Eq.(\ref{eq:dictionary}), this requires a knowledge of all three neutrino masses.  We plot in Fig.1 the ratios of the eigenvalues as a function of $m_2/m_1$, assuming TBM mixing, normal hierarchy, and from the data, $\Delta m_{31}^2 = 32\Delta m_{21}^2$. 

\begin{figure}[h!] \label{fig:eigenvalues}
 \centering
\scalebox{0.8}{\includegraphics*{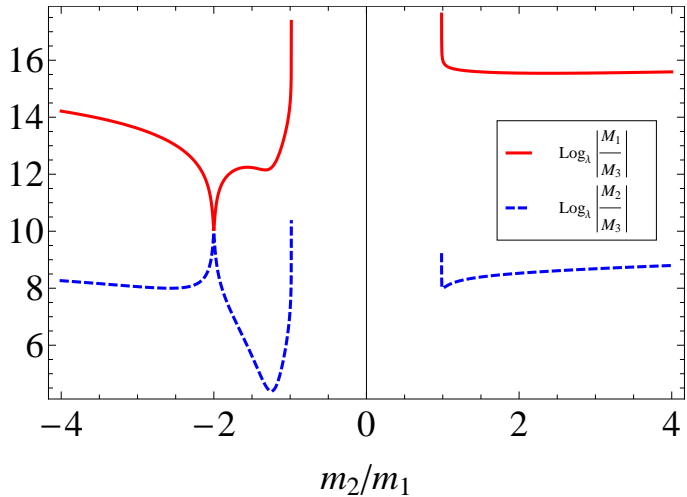}} 
\caption{A plot of the logarithmic ratios of eigenvalues of $\m M$ as a function of the neutrino mass ratio $m_2 /m_1$.}
\end{figure}

We observe, when $m_2/m_1=-2$,  a kink at which two eigenvalues are nearly degenerate, and the hierarchy between the lightest and heaviest mass is less severe. No such kink exists for the inverted hierarchy.

We now study this singular case, and how it can arise naturally in $\bs{\m Z_7 \rtimes \m Z_3}$.  

%%%%%%%%%%%%%%%%%%%%%%%%
\section{A Special Majorana Matrix}

When $m_2/m_1=-2$ and $\tan^2\eta^{}_{12}=1/2$, $a^{}_{22}$ and $a^{}_{23}$ are related by

\begin{eqnarray}
a^{}_{22} ~=~ -a^{}_{23} ~=~ \frac{1}{2 m_3}.
\end{eqnarray}
This simple relation is the key to understanding the kink:  it makes the sub-determinant of $\m M$'s (23) block  vanish, and yields near degeneracy of its two lightest eigenvalues. 

More generally, the relation $a^{}_{22}=-a^{}_{23}$ applied to a seesaw mixing matrix with $\eta^{}_{23}=45^\circ$ and $ \eta^{}_{13}=0^\circ$ yields a Gatto-Sartori-Tonin type  relation\cite{Go_Gatto!},
 
\begin{equation}
\label{eq:prediction}
\frac{m_1}{m_2} = -\tan^2\eta_{12},
\end{equation}
where $\eta_{12} = \theta_{12}$ modulo Cabibbo effects.\footnote{Using the best fit value for $\theta_{12}= 33.6^\circ$, the sizes of Cabibbo effects for TBM and GR mixings are around  $\lambda^2/2$ and $-2\lambda^2/3$ respectively.} In particular, for GR mixing, a similar kink appears at $m_2/m_1=-(1+\sqrt{5})/2 \approx -1.6$.  

For TBM mixing, Eq.(\ref{eq:prediction}) predicts 

\be
\label{eq:TBMprediction}
m_1 \sim 0.005 ~\textrm{eV} , \qquad m_2 \sim 0.01 ~\textrm{eV}, \qquad m_3 \sim 0.05 ~\textrm{eV},
\ee
while for GR mixing, we have 

\be
\label{eq:GRprediction}
m_1 \sim 0.0068 ~\textrm{eV} , \qquad m_2 \sim 0.011 ~\textrm{eV}, \qquad m_3 \sim 0.05 ~\textrm{eV}.
\ee
Note that BM mixing is not compatible with Eq.(\ref{eq:prediction}), as it gives an exact degeneracy of $m_1$ and $m_2$, and therefore no oscillation.

From now on we focus on TBM mixing, where the Majorana matrix assumes the elegant form, 

\begin{eqnarray} 
{\m M } = \begin{pmatrix}
{r \lambda^{16} & r \lambda^{12} & r \lambda^{8} \cr r \lambda^{12} & \lambda^{8} & -\lambda^{4} \cr r \lambda^{8} & -\lambda^{4} & 1  
}\end{pmatrix},\qquad r~=~\frac{m_3}{m_1}.
\end{eqnarray} 
The data requires $r \approx 10$, which is not so large as to affect the hierarchy. Up to an overall normalization, the spectrum of $\m M$ is given by

\begin{eqnarray} \nonumber 
|M_3| &=& 1 + \lambda^8 + \m O(\lambda^{16}), \\ \nonumber 
|M_1| + |M_2|  &=& 4 r \lambda^{12}  +  \m O(\lambda^{16}), \\ 
\frac{|M_2| - |M_1|}{|M_1| + |M_2|} &=& \frac{1}{4} ( r - 1 )  \lambda^4 +  \m O(\lambda^{12}),
\end{eqnarray}
which shows a single large $r\lambda^{12}$ hierarchy with two lighter and almost degenerate Majorana neutrinos. When the largest eigenvalue of $\m M$ corresponds to the GUT scale of $\sim 10^{16}$ GeV, this gives $M_1 \approx M_2 \approx 10^9$ GeV.  Such a degeneracy may be physically appealing from the point of view of leptogenesis, placing such a model in the case of ``resonant leptogenesis"\cite{resonant}.

This special matrix appears to be extremely fine-tuned, containing a vanishing sub-determinant. However, this determinant can naturally arise  in higher-dimensional invariants.\footnote{We thank T. Kephart  for pointing us in the right direction.} We therefore turn to the question of building this matrix from  higher dimensional operators.  Indeed, we find that our matrix  can arise naturally from the dimension-five and dimension-six invariants of ${\bs{\m Z_7\rtimes\m Z_3}}$. 

%%%%%%%%%%%%%%%%%%%%%%%%%%%%%%%%%%%

\section{Building the Special Matrix in $\bs{\m Z_7\rtimes\m Z_3}$}

We construct $\m M$ by coupling the right-handed neutrinos, $N$ to additional ``familon'' fields $\varphi$ and $\bar{\varphi}$ which are gauge singlets, family triplets and anti-triplets. The hierarchy in the special matrix is assumed to be generated by familon vacuum values, so that the $(\bs{\m Z_7\rtimes\m Z_3})$-invariant operators which generate $\m M$ are of the same dimension
%, with the same order of couplings if more than one operator is involved
.\footnote{This is in contrast to the Frogatt-Nielsen scheme\cite{FN}, in which the hierarchy is generated by assembling operators with different dimensions.} 

The special matrix involves precise relations among its elements. They can be generated by linear combinations of $(\bs{\m Z_7\rtimes\m Z_3})$-invariant operators, or better by a single operator.  In the first case,  additional symmetries are required to determine the linear combination; the single operator case arises at dimension-six.  

Both cases are achieved with higher-dimension couplings, thought to arise from ``integrating out"  heavier field(s). Besides being invariant under the family symmetry, they have a  {\em nesting} structure, corresponding to a particular linear combination of invariants.  

Nesting then suggests  an underlying renormalizable structure which corresponds to a particular way of building  invariants by combining particular covariants. By ``{\em  nesting}'' we mean a particular choice of covariants. For dimension-five and -six  interactions, we consider all possible nestings, 

$$ 
((A\otimes B)\otimes(C\otimes D)),\qquad  (((A\otimes B)\otimes(C\otimes D))\otimes E),$$
where the inner parentheses single out a given covariant, and  $A,B,C,D,E,$ are any  permutations of two $N$'s and familons. Considering all posssible nestings gives an over-complete list of the possible matrices for a given operator dimension.  They are tabulated in  Appendix A up to dimension-five and dimension-six operators. The family symmetry determines all possible linearly independent invariants, and a nesting singles out a \emph{specific} linear combination. As any bird knows, {\em nesting matters!}

Nestings that  produce the special matrix are found by careful inspection of their structures. The two examples mentioned above are indeed obtained in this way.

\subsection{Two Dimension-five $(\bs{\m Z_7\rtimes\m Z_3})$-Invariant Operators}

At dimension five, at least two $\bs{\m Z_7\rtimes\m Z_3}$-invariant operators are needed to produce the special matrix. 

For example, a linear combination involving two anti-triplet familons generates the Majorana matrix,

\begin{eqnarray}
\label{eq:dim5}
\begin{pmatrix}{
2 \bar{\varphi}_1\bar{\varphi}_1^\prime & -(\bar{\varphi}_1\bar{\varphi}_2^\prime + \bar{\varphi}_2\bar{\varphi}_1^\prime) & -(\bar{\varphi}_1\bar{\varphi}_3^\prime + \bar{\varphi}_3\bar{\varphi}_1^\prime) \cr
 & 2 \bar{\varphi}_2\bar{\varphi}_2^\prime & -(\bar{\varphi}_2\bar{\varphi}_3^\prime + \bar{\varphi}_3\bar{\varphi}_2^\prime) \cr
 & & 2\bar{\varphi}_3\bar{\varphi}_3^\prime
}\end{pmatrix}.
\end{eqnarray}
Counting the number of independent parameters, we derive a constraint from the above matrix among the prefactors of $\m M$ which leaves $r$ undetermined, 

\begin{equation}
(a_{12}a_{33} + a_{13} a_{23})^2 = (a^2_{13} - a_{11}a_{33})(a^2_{23} - a_{22}a_{33}).
\end{equation}
If one imposes $\eta_{23} = 45^\circ$ and $\eta_{13} = 0^\circ$, that is $a_{12} = a_{13}$ and $a_{22} = a_{33}$, it reduces to

\begin{equation} \label{constraint}
a_{12}^2(a_{22} + a_{23})^2 = (a^2_{12} - a_{11}a_{33})(a^2_{23} - a^2_{33}).
\end{equation}
with solutions along two branches:

$$\cases{a_{22} = - a_{23}, \cr a_{22} \neq -a_{23}.}$$
The first condition $a_{22} = - a_{23}$ is precisely the extra relation that gives the special matrix. 

The familon vacuum values are, 

\begin{equation} \label{eq:famvac1}
\bar{\varphi} = \langle \bar\varphi_1 \rangle \begin{pmatrix}{
\bar{\alpha} \lambda^8 \cr  \lambda^4 \cr 1
}\end{pmatrix}, \qquad \bar{\varphi}^\prime= \langle \bar\varphi_1' \rangle  \begin{pmatrix}{
\bar{\alpha}^\prime \lambda^8 \cr \lambda^4 \cr 1
}\end{pmatrix},
\end{equation}
along with the additional relation,

\begin{eqnarray} 
\bar{\alpha}\bar{\alpha}^\prime = - \frac{1}{2} (\bar{\alpha}+\bar{\alpha}^\prime) = r.
\end{eqnarray}

Any linear combination, with precise relations amongst different operators, points to additional symmetries. In a future publication, we explore how such linear combinations can arise naturally by enlarging the symmetry group from $\bs{\m Z_7\rtimes\m Z_3}$.

\subsection{Single Dimension-six $(\bs{\m Z_7\rtimes\m Z_3})$-invariant Operator}

A single operator that produce the special matrix first arises at dimension six. The dimension-six $({\bs{\m Z_7\rtimes\m Z_3}})$-invariant couplings are of four types,

$$NN\cases{\varphi\,\varphi'\,\varphi''\cr \varphi\,\varphi'\,\overline\varphi\cr
 \varphi\,\overline\varphi\,\overline\varphi'\cr\overline\varphi\,\overline\varphi'\,\overline\varphi'' }.$$
One finds  three linearly independent invariants of the first type, five of the second and third types , and three of the fourth type. All nestings generate a long list (seven pages!) of possible linear combinations of these invariants. 

We find several nestings capable of reproducing the special matrix. With details to be found in Appendix B, they are grouped by structure into three classes: 

\begin{itemize}

\item In the first class of invariants, each matrix element is a monomial in the familon fields, and all satisfy a constraint amongst the prefactors $a_{ij}$. All are 
  {\em over}-constrained,  and lead to $r = \pm (1/8)$, incompatible with the data. 

\item In the second class, the matrix elements are combinations of monomials. We can obtain the special Majorana matrix in two ways. In the first, one vacuum value of the familon components vanishes, and in the other {\em none } of the familon components vanish in the vacuum. Both cases are {\em under}-constrained and do not impose constraints on the pre-factors. 

\item The third class contains only one nesting! It is neither over nor under constrained, and singles out the special matrix. It arises from the unique nesting, 

\begin{eqnarray}
((N \varphi)_{\bf{\bar{3}}_+} (N \varphi^\prime)_{\bf{\bar{3}}_+})_{\bf{3}_-} \bar{\varphi},
\end{eqnarray}
which generates the  Majorana matrix,

\be \begin{pmatrix}{
2 \bar{\varphi}_1 B_{23} & \bar{\varphi}_1 B_{13} - \bar{\varphi}_2 B_{23} & -\bar{\varphi}_1 B_{12} - \bar{\varphi}_3 B_{23}  \cr & -2 \bar{\varphi}_2 B_{13} & -\bar{\varphi}_2 B_{12} + \bar{\varphi}_3 B_{13}  \cr & & 2 \bar{\varphi}_3 B_{12}
}\end{pmatrix},
\ee
where 
$$
B_{ij} = \varphi_i \varphi^\prime_j - \varphi_i^\prime \varphi_j. $$  

Interestingly, one notices that the above matrix has the same form as that of Eq.(\ref{eq:dim5}) by letting 
\be
\bar{\varphi}_1^\prime = B_{23}, \qquad 
\bar{\varphi}_2^\prime = B_{31},  \qquad
\bar{\varphi}_3^\prime = B_{12},
\ee
so that the constraint derived previously also holds here.

The familon vacuum values are ($\varphi \neq \varphi^\prime$), 

\begin{equation} \label{eq:famhierarchy}
\bar{\varphi} \sim \begin{pmatrix}{
\bar{\alpha} \lambda^8 \cr  \lambda^4 \cr 1
}\end{pmatrix}, \qquad \varphi \sim \begin{pmatrix}{ 1 \cr \alpha \lambda^4 \cr \beta \lambda^8 }\end{pmatrix} , \qquad \varphi^\prime \sim \begin{pmatrix}{ 1 \cr \alpha^\prime \lambda^4 \cr \beta^\prime \lambda^8 }\end{pmatrix},
\end{equation}
along with the additional relations,

\begin{eqnarray} 
\frac{\bar{\alpha}^2}{1+ 2\bar{\alpha}} ~=~\bar{\alpha} (\alpha + \beta) ~=~ \bar{\alpha}(\alpha^\prime + \beta^\prime) ~=~ - r .
\end{eqnarray}

\end{itemize}

The form of this particular nesting, $((N \varphi)_{\bf{\bar{3}}_+} (N \varphi^\prime)_{\bf{\bar{3}}_+})_{\bf{3}_-} \bar{\varphi}$, suggests an underlying theory with two family triplet familon fields $F$ and $F'$. The underlying superpotential follows

\be
W_u'= a(N\Phi)_{\bf \bar 3_+} F+g(\bar F\bar F)_{\bf 3_-} \bar\varphi+M^{}_FF\bar F,
\ee
where we have assigned an extra $\bs{\m S_3}$ symmetry, with $\Phi = (\varphi,\varphi')$ and $F$ as $\bs{\m S_3}$ doublets. The coupling of the $\bs{\m S_3}$-invariant combination $\bar F\bar F$ requires $\bar\varphi$  to transform as a $\bf 1'$ under $\m S_3$. With the same parity as in the dimension-five case, it yields the group structure found in the dimension-five case, albeit with different symmetry assignments for the underlying particles. 

In this case,  supersymmetry is always required to explain  the absence of the $(N\Phi)_{\bf \bar 3_-}F$ coupling. In addition, one finds that several  dimension-four couplings are still allowed by $(\bs{\m Z_7\rtimes\m Z_3})\times \bs{\m  S_3}\times \bs{\m Z_2}$.  To explain their absence, one may have to appeal to supersymmetry, or require additional symmetries. This can be done by assigning to the fields an addition $\bs{\m Z_3}\times \bs{\m Z_2'}$ symmetry. Clearly,  this singling out of the coupling we require to produce the magic matrix is much more elaborate than in the dimension-five case. 

Although these two cases are quite distinct, hierarchies arise from the familon vacuum values, yet to be determined from the hitherto unknown potential. These same familons  may  couple to the charge $(2/3)$ quarks. We next show that the pattern of hierarchies of Eq.(\ref{eq:famhierarchy}) obtained in producing the special matrix is compatible with the charge $(2/3)$ sector, and how this sector can  provide partial information on the \emph{absolute} hierarchy of the familons.

%%%%%%%%%%%%%%%%%%%%%%%%%%%%%%%%%%%%%%%%%%%%

\section{Charge (2/3) sector}

The Majorana sector contains some information on familon structures; we next want to see if these same structures are compatible with model building in the charge (2/3) sector. 

\begin{itemize}

\item When the Majorana matrix is generated from dimension-five invariants, two anti-triplet familons $\bar{\varphi}$ and $\bar{\varphi}^\prime$ are introduced. Beginning from the superpotential of Eq.(\ref{eq:potential}), the dimension-four Yukawa coupling $(\chi\chi)\m H_u$ will generate the top quark mass when $\m H_u$ acquires the vacuum value of Eq.(\ref{huvev}), leaving the up and charm quarks massless,

$$
W = y^{}_{10} \chi^{}_3 \chi^{}_3 v^{}_u + \ldots $$

As in the Majorana sector, we assume the remaining hierarchy is filled by higher-dimensional ($\bs{\m Z_7 \rtimes \m Z_3}$)-invariant interactions involving familon fields. 

\begin{itemize}
\item With only $\bar{\varphi}$ and $\bar{\varphi}^\prime$, we find no dimension-five or dimension-six invariants capable of generating the remaining hierarchy. In this case we are led to assume that it is carried by the Higgs particles with the vacuum value,

\begin{eqnarray}
\langle \m H_u \rangle = v_u \begin{pmatrix}{
\lambda^8 \cr \lambda^4 \cr 1
}\end{pmatrix},
\end{eqnarray}
so that the tree-level interaction $(\chi\chi)\m H_u$ will generate the full hierarchy of the charge (2/3) sector. 

\item Alternatively, one can try adding additional familons while keeping the vaccum value of $\m H_u$ to be the same as that in Eq.(\ref{huvev}). The simplest possibility is to enlarge the familon sector to include another triplet familon field $\varphi$. Besides the tree-level interaction $(\chi\chi)\m H_u$, which generates the top quark mass, one then needs one of the following dimension-six invariants

\begin{eqnarray}
\left[ (\chi\chi)^{}_{\bf{3}} ~ (\m H^{}_u\bar{\varphi})_{\bf{3_+}} \right]_{\bf{\bar{3}_\pm}} \varphi &\longrightarrow & \frac{v_u^{}}{2\sqrt{3}} \begin{pmatrix}{
\pm\varphi^{}_1\bar{\varphi}^{}_2 & & \cr
& \varphi^{}_1\bar{\varphi}^{}_3 & \cr
& & \varphi^{}_2\bar{\varphi}^{}_2 \pm \varphi^{}_3\bar{\varphi}^{}_3
}\end{pmatrix}, \nonumber \\
\left[ (\chi\chi)^{}_{\bf{3}} ~ (\m H^{}_u\bar{\varphi})^{}_{\bf{3_-}} \right]_{\bf{\bar{3}_\pm}} \varphi &\longrightarrow & \frac{v_u^{}}{2\sqrt{3}} \begin{pmatrix}{
\mp\varphi^{}_1\bar{\varphi}^{}_2 & & \cr
& \varphi^{}_1\bar{\varphi}^{}_3 & \cr
& & -\varphi^{}_2\bar{\varphi}^{}_2 \pm \varphi^{}_3\bar{\varphi}^{}_3
}\end{pmatrix}, \nonumber
\end{eqnarray}
to fulfill the remaining hierarchy.\footnote{If $\bar \phi$ is a component of an $\m S_3$ doublet, then $\varphi$ is required to be part of an $\m S_3$ doublet as well, $\Phi = (\varphi, \varphi^\prime)$, and there will be an additional term involving $\varphi^\prime$. However, this additional term will give the same structure and will not affect Eq.(\ref{eq:massratio1}).}

Taking the first symmetric combination as an example, the superpotential in this case is,

\begin{eqnarray} \label{eq:example2}
W &=& y^{}_{10} \chi \chi \m H_u + \frac{y_{10}^\prime}{M_X^2} \left[ (\chi\chi)_{\bf{3}} ~ (\m H^{}_u\bar{\varphi})_{\bf{3_+}} \right]_{\bf{\bar{3}_+}}\varphi + \ldots \\ \nonumber 
&\rightarrow&  y^{}_{10} \chi^{}_3 \chi^{}_3 v_u +   \frac{y^\prime_{10}\langle \varphi^{}_1 \rangle \langle \bar{\varphi}^{}_3 \rangle}{2\sqrt{3}M_X^2}\left( \lambda_{}^4 \chi^{}_1 \chi^{}_1 +  \chi^{}_2 \chi^{}_2 \right) v^{}_u + \ldots, 
\end{eqnarray}
yielding 

\begin{equation} \label{eq:massratio1}
\frac{m_u^{}}{m_c^{}}=\frac{\langle \bar{\varphi}^{}_2 \rangle}{ \langle \bar{\varphi}^{}_3 \rangle }= \lambda_{}^4,
\end{equation}
the correct mass hierarchy between the first two families. 

However, this nesting requires a tuning of couplings to explain $m_c / m_t$, 

\be \label{eq:massratio2}
\frac{1}{2 \sqrt{3}} \frac{y^\prime_{10}}{y^{}_{10}} \frac{\langle \varphi^{}_1 \rangle \langle \bar{\varphi}^{}_3 \rangle}{M_X^2} = \lambda_{}^4 .
\ee
Note that such a relation constrains the absolute magnitude of the familon vacuum values, a feature unavailable by simply considering the Majorana sector. 

Although both nestings introduce non-zero contributions to the third family, they are suppressed compared to the tree-level coupling.

\end{itemize}

\item If $\m M$ is instead produced by a single dimension-six invariant, we must have two triplet familons $\varphi$ and $\varphi^\prime$, and one anti-triplet familon $\bar{\varphi}$. If the vacuum value of $\m H_u$ is still assumed to be that of Eq.(\ref{huvev}), higher-dimensional ($\bs{\m Z_7 \rtimes \m Z_3}$)-invariants are also required to produce the remaining hierarchy. We find two ways to achieve this goal. 

\begin{itemize}

\item Since the superpotential of Eq.(\ref{eq:example2}) involves $\varphi$ and $\bar \varphi$, and the possible vacuum structures of the anti-triplets given by Eq.(\ref{eq:famvac1}) and Eq.(\ref{eq:famhierarchy}) are identical, we may use the same superpotential to generate the correct mass hierarchy in this case; the relations of Eq.(\ref{eq:massratio1}) and Eq.(\ref{eq:massratio2}) then follow identically.

\item Instead one may use invariants with two triplet familons $\varphi$ and $\varphi^\prime$ in the following nestings

\begin{eqnarray}
\left[ (\chi\chi)_{\bf{3}} ~ (\m H_u\varphi)_{\bf{3}} \right]_{\bf{\bar{3}_\pm}} \varphi^\prime \longrightarrow  \frac{v_u^{}}{\sqrt{6}} \begin{pmatrix}{
\varphi_2\varphi_3^\prime & & \cr
& \pm \varphi_2\varphi_2^\prime & \cr
& & 0}\end{pmatrix}. \nonumber
\end{eqnarray}
Taking the symmetric combination, the superpotential would then be, 

\begin{eqnarray} 
W &=& y^{}_{10} \chi \chi \m H^{}_u +  \frac{y_{10}^\prime}{M_X^2} \left[ (\chi\chi)^{}_{\bf{3}} ~ (\m H^{}_u\varphi)_{\bf{3}} \right]_{\bf{\bar{3}_+}} \varphi^\prime + \ldots \\ \nonumber
&\rightarrow&  y^{}_{10} \chi^{}_3 \chi^{}_3 v^{}_u +  \frac{y^\prime_{10}\langle \varphi^{}_2 \rangle \langle \varphi_2^\prime \rangle}{\sqrt{6}M_X^2}\left( \frac{\beta^\prime}{\alpha^\prime} \lambda^4 \chi^{}_1 \chi^{}_1 +  \chi^{}_2 \chi^{}_2 \right) v^{}_u + \ldots, 
\end{eqnarray}  
where $y^\prime_{10}$ is a dimensionless coupling constant and $M_X$ is an unknown heavy scale.  We see that the correct hierarchy between the first and second flavors can be reproduced with the additional constraint $\alpha^\prime = \beta^\prime$.

The additional $\lambda^4$ factor present between the third and the first two flavors implies the further tuning,

\be 
 \frac{1}{\sqrt{6}}\frac{y^\prime_{10}}{y^{}_{10}}\frac{\langle \varphi^{}_2 \rangle \langle \varphi_2^\prime \rangle}{M_X^2} = \lambda_{}^4 ,
\ee
this time involving the absolute scale of $\varphi$ and $\varphi^\prime$.
\end{itemize}
\end{itemize}

Interestingly, utilizing the same familons of the Majorana sector in the charge $(2/3)$ sector can give complementary information on the unknown physics of the familon sector; one gives constraints on the relative ratio of their components, the other on their absolute scale. In addition, although the ratio $m_c / m_t$ remains to be explained, this sharing of familons between the up-quark Yukawa matrix and the Majorana matrix may partially explain why the same hierarchical factor $\lambda^4$ can appear in both sectors.

%%%%%%%%%%%%%%%%%%%%%%%%%%%%%%%%%%%%%%%%%%%%%%%%%%%%%%%%%%%%%%%%%%%%%%%%%%%%%%%%%%%%%

\section{Summary and Conclusions}
In this paper, we have investigated the addition of a discrete family symmetry to the SSM.  The quark and charged lepton mass matrices display a hierarchical structure not seen in the neutrino sector.  Within the context of grand unification, this is a bit of a mystery; the neutral Dirac Yukawa matrix $Y^{(0)}$ is expected to have the same structure as the up-quark Yukawa matrix $Y^{(2/3)}$.  In this work we have addressed this issue within the context of a $\bs{\m Z_7\rtimes\m Z_3}$ family symmetry.

Taking $Y^{(2/3)}$ strongly hierarchical and diagonal, and assuming $Y^{(0)}\sim Y^{(2/3)}$, we exploit the seesaw mechanism to reproduce the observed neutrino mass pattern.  However, in order to compensate for the strong hierarchy in $Y^{(0)}$, the associated high-scale Majorana mass matrix itself must show a strong, specific hierarchical structure, which may indicate a deep relation between the Yukawa and Majorana matrices.  Additionally, measurement of the neutrino mass and mixing parameters leads to additional desired features of the Majorana matrix.  

We  chose a special Majorana matrix which is strongly hierarchical and consistent with Tri-bimaximal mixing.  This special matrix predicts the normal hierarchy and yields a testable relation between light neutrino masses and the solar mixing angle
\begin{equation}
\frac{m_1}{m_2} = -\tan^2\eta_{12}.
\end{equation}
Furthermore, the special matrix gives the heavy right-handed neutrinos reasonable masses; in future work, we intend to investigate whether these values are compatible with leptogenesis. In particular, although the approximate degeneracy of the right-handed neutrino spectrum looks appealing from the point of view of ``resonant leptogenesis'', careful attention to the values of CP phases must be given in order to ensure that leptogenesis is successful.  

We take the hierarchy in the special matrix to be produced not by a hierarchy of couplings, but instead by the structure of the vacuum values of familon fields.  We thus look for specific operators invariant under $\bs{\m Z_7\rtimes\m Z_3}$ which can produce the special matrix for some familon vacuum values. Our search suggests either a linear combination of several dimension-five operators or one dimension-six operator. They contain two or three familon fields, respectively, single out the special matrix, and are compatible with the light neutrino masses.  

The same familon fields can be used to construct the terms responsible for the up and charm quark masses in $Y^{(2/3)}$, giving constraints on the magnitude of the familon fields. Thus, with an appropriate familon sector, the $\bs{\m Z_7\rtimes\m Z_3}$ family symmetry can both reproduce some of the hierarchy in the up-quark sector and generate the Majorana mass terms needed to largely erase the hierarchy in the physical light neutrino masses.
 
Our aim in this publication has been to reduce the number of familon couplings to the Standard Model, and from there infer the number of familons needed and their  symmetries. We leave the question of the familon scalar potential for a more complete model; at this stage we believe such a question to be premature. Familons themselves may be effective fields, coming perhaps from a more fundamental (extra dimensions, branes?) theory. However, we believe the vacuum structure of the familon sector will yield useful information, pointing the way to a deeper theory.

A complete description of the neutrino data and the lepton sector requires specification of the charged-lepton Dirac matrix $Y^{(-1)}$.  Its diagonalization provides not only the charged-lepton masses, but also corrections to the neutrino mixing angles away from their seesaw values. Such corrections are necessary for all three popular mixing schemes (TBM, GR, and BM) discussed in the text, and implementing a suitable pattern of corrections is nontrivial but not impossible. The interested reader may find a specific example of how this can be accomplished, using ideas from grand unification, in Appendix C. In a forthcoming paper \cite{nextpaper}, we plan to further study the question of producing $Y^{(-1)}$ compatible with the neutrino mass and mixing parameters; from this, one can make predictions for leptonic CP violation.      

In summary, we find that the $\bs{\m Z_7\rtimes\m Z_3}$ family symmetry appears very encouraging for producing fermion mass parameters compatible with observation, and it shows promise for additional related studies: on the  familon physics implied by the higher dimension operators; on a possible consistency with resonant leptogenesis; and, finally, on the implications of the assignment of family quantum numbers to the Higgs bosons.

%%%%%%%%%%%%%%%%%%%%%%%%%%%%%%%%%%%%%%%
%%%%%%%%%%%%%%%%%%%%%%%%%%%%%%%%%%%%

\section{Acknowledgements}

We thank James Gainer for his useful discussions. One of us (MP) would like to thank the McKnight Doctoral Fellowship Program for their continued support. One of us (PR) also wishes to thank the Aspen Center for Physics, where part of this work was performed. This research is partially supported by the Department of Energy Grant No. DE-FG02-97ER41029.

%%%%%%%%%%%%%%%%%%%%%%%%%%%%%%%%%%%%%%%

\newpage
%%%%%%%%%%%%%%%%%%%%%%%%%%%%%%%%%
\noindent{\bf \Large Appendix A: ${\bs{\m Z_7\rtimes\m Z_3}}$}
\vskip .5cm
\noindent 
The Frobenius group of order $21$ is the smallest finite non-Abelian subgroup of $SU(3)$. It contains elements of order three and seven, with the presentation  $\langle c\, , d\,|\,  c^7=d^3=1\, , \,d^{-1}cd=c^4\,\rangle$. 

Its irreducible representations are, a real singlet, one complex triplet $\bf 3$, a complex singlet, $\bf 1'$, and their inequivalent conjugates, $\bf \overline  3$, and   $\bf\overline  1'$. Their Kronecker products are ($\pm$ refers to symmetry/antisymmetry)
 
 \bean
 {\bf 1'}\otimes {\bf 1'}&=&{\bf \overline 1'}, \qquad\qquad\qquad ~~~ 
 {\bf 1'}\otimes {\bf \overline1'}~=~{\bf 1}\\
{\bf 3}\:\otimes\!\:{\bf 1'}&=&{\bf 3},\qquad\qquad\qquad~~~ ~
 {\bf 3}\:\otimes\!\:{\bf \overline 1'}~=~{\bf 3}\\ 
 {\bf 3}\,\otimes\,{\bf 3}\:&=&({\bf 3}+{\bf\overline 3})^{}_++{\bf \overline 3}_-^{},\qquad  
 {\bf 3}\,\otimes\,{\bf \overline 3}~=~{\bf 1}+{\bf 1'}+{\bf \overline 1'}+{\bf 3}+{\bf \overline3}. 
\eean
with Clebsch-Gordan decompositions,

\bean
({\bf 3}\otimes {\bf 3})_+^{}~&\longrightarrow&~~{\bf 3}:~~\cases{\ket 3\ket{3'}\cr \ket 1\ket{1'}\cr\ket 2\ket{2'}}\ ;\qquad ~~\longrightarrow~~{\bf\overline 3}:~~\cases{\frac{1}{\sqrt{2}}\left(\ket 3\ket{2'}+~\ket2\ket{3'}\right)\cr\frac{1}{\sqrt{2}}\left(\ket 1\ket{3'}+~\ket 3\ket{1'}\right)\cr\frac{1}{\sqrt{2}}\left(\ket 2\ket{1'}+~\ket1\ket{2'}\right)}\;\\
({\bf 3}\otimes {\bf 3})_-^{}~&\longrightarrow&~~ {\bf\overline 3}:~~\cases{\frac{1}{\sqrt{2}}\left(\ket 3\ket{2'}-~\ket2\ket{3'}\right)\cr\frac{1}{\sqrt{2}}\left(\ket 1\ket{3'}-~\ket 3\ket{1'}\right)\cr\frac{1}{\sqrt{2}}\left(\ket 2\ket{1'}-~\ket1\ket{2'}\right)}\ .
\eean

\bean
{\bf 3}\otimes {\bf\overline 3}~&\longrightarrow~~&{\bf 3}:~~\cases{\ket 2\ket{\overline  1}\cr \ket 3\ket{\overline  2}\cr\ket 1\ket{\overline  3}}\ ;\qquad\qquad ~\longrightarrow~~{\bf\overline 3}:~~\cases{\ket 1\ket{\overline  2}\cr \ket 2\ket{\overline  3}\cr\ket 3\ket{\overline  1}}\ ,\\
&&\\
%\eean
%\bean
{\bf 3}\otimes {\bf\overline 3}~&\longrightarrow~~&{\bf 1{\phantom{'}}}:~~\frac{1}{\sqrt{3}}\,\left( \ket 1\ket{\overline  1}+ ~\ket
  2\ket{\overline  2}+ ~ \ket 3\ket{\overline  3}\right)\ , \nonumber \\ 
{\bf 3}\otimes {\bf\overline 3}~&\longrightarrow~~&{\bf 1'}:~~\frac{1}{\sqrt{3}}\,\left( \ket 1\ket{\overline  1}+ ~\omega^2_{}\ket
 2\ket{\overline  2}+ ~\omega\:\ket 3\ket{\overline  3}\right)\ ,~~~~~~~~~~~~~\\
{\bf 3}\otimes {\bf\overline 3}~&\longrightarrow~~&{\bf \overline 1'}:~~\frac{1}{\sqrt{3}}\,\left( \ket 1\ket{\overline  1}+ ~\omega\:\ket  2\ket{\overline  2}+~ \omega^2_{}\ket 3\ket{\overline  3}\right)\ ,~~\omega=\exp(2i\pi/3) \\ \\ {\bf 1^\prime}\otimes{\bf 3}~&\longrightarrow&~~ {\bf 3}:~~\cases{s^\prime \ket 1 \cr s^\prime \omega \ket 2 \cr s^\prime \omega^2 \ket 3} \\ 
{\bf {\overline 1}^\prime}\otimes{\bf 3}~&\longrightarrow&~~ {\bf 3}:~~\cases{{\overline s}^\prime \ket 1 \cr {\overline s}^\prime \omega^2 \ket 2 \cr {\overline s}^\prime \omega \ket 3} \nonumber
\eean

%A similar but much longer list for dimension-six Majorana invariants has been determined; it is much too long to appear here, and is available in Mathematica.
%%%%%%%%%%%%%%%%
\newpage
%%%%%%%%%%%%%%%%
 
\noindent{ \bf \Large${\bs{\m Z_7\rtimes\m Z_3}}$ Invariants}
\vskip .5cm
   
\noindent In our model, invariants are constructed out of ${\bs{\m Z_7\rtimes\m Z_3}}$-triplet matter fields, $\psi$, $\chi$, and $N$,  family antitriplets Higgs fields, $\m H_{u,d}$, and familons which can be either family triplets, $\varphi$,  or antitriplets, $\bar\varphi$. 

Since we consider in this paper only the Majorana and charge (2/3) sectors, both of which produce symmetric matrices, we display here only the Majorana invariants. Invariants for the charge (2/3) sector can then be easily obtained by replacing $N N$ by $\chi \chi$. 

%$$
%{\m H}^{}_{u,d}=\begin{pmatrix}{H^{}_{u,d1}\cr H^{}_{u,d2}\cr H^{}_{u,d3}}\end{pmatrix}~\sim~ {\bf \overline  3},\qquad 
%\varphi=\begin{pmatrix}{\varphi^{}_1\cr \varphi^{}_2\cr \varphi^{}_3}\end{pmatrix}\sim {\bf 3},\quad
 %\overline \varphi=\begin{pmatrix}{\overline \varphi^{}_1\cr \overline \varphi^{}_2\cr \overline \varphi^{}_3}\end{pmatrix}\sim {\bf \overline  3}.$$
\vskip .3cm
\noindent {\large \bf Dimension-four Majorana invariants}
\vskip .3cm
\noindent The tree-level couplings are all diagonal.

$$(NN)^{}_{\bf 3}H^{}_u:~~~ \longrightarrow~~~N^{}_{}\begin{pmatrix}{ H^{}_{u2}&0&0\cr 0&H^{}_{u3}&0\cr 0& 0&H^{}_{u1}}\end{pmatrix}N
.$$

%%%%%%%%%%%%%%%%%%%%%%%%%%%%%%%%
\vskip .3cm
\noindent{\large\bf Dimension-five Majorana invariants}
\vskip .3cm
\noindent They are of the form, 

$$ NN\cases{\varphi\varphi'\cr \varphi\overline\varphi\cr {\overline \varphi}~ {\overline \varphi}^\prime},$$
with different nesting schemes. 

As mentioned in the text, we have chosen to display all invariants with a particular nesting scheme,
$$ 
((~~\otimes~~)\otimes(~~\otimes~~))\otimes~~).$$ 
However, one may have also chosen a different nesting scheme, such as

$$
(~~\otimes (~~\otimes (~~\otimes (~~\otimes ~~)))).$$
Fortunately, for $\bs{\m Z_7 \rtimes \m Z_3}$, we have found that considering all possible permutations of the fields within both nesting schemes will give the same set of matrices, rendering them equivalent. Since, the second generates a much longer list of possibilities, we have displayed all invariants using the first nesting scheme. 

We organize the list in terms of the familons, and choose to express a particular nesting in terms of linear combinations of particular simple linearly independent invariants.  In the following, independent invariants involving different sets of familons are distinguished by distinct letters, while the superscripts denote the nesting order of the familon fields. 
\vskip .3cm

\newpage
\begin{itemize}
\item There are two types of linearly independent $NN\varphi\varphi^\prime$ invariants:

\begin{eqnarray}
\m I_1^{(\varphi,\varphi^\prime)} &=& \frac{1}{2} \left( (NN)_{\bf{3}}~ (\varphi\varphi^\prime)_{\bf{\bar{3}_+}} + (NN)_{\bf{3}}~ (\varphi\varphi^\prime)_{\bf{\bar{3}_-}} \right ) \nonumber \\
&\longrightarrow & \frac{1}{\sqrt{6}}
\begin{pmatrix}{
\varphi_1\varphi_3^\prime & 0 & 0 \cr
& \varphi_2\varphi_1^\prime & 0 \cr
& & \varphi_3\varphi_2^\prime
}\end{pmatrix}, 
\nonumber \\
\m I_2^{(\varphi,\varphi^\prime)} &=& (NN)_{\bf{\bar{3}}}~ (\varphi\varphi^\prime)_{\bf{3}} \nonumber \\ 
&\longrightarrow & \frac{1}{\sqrt{6}}
\begin{pmatrix}{
0 & \varphi_2\varphi_2^\prime & \varphi_1\varphi_1^\prime \cr
& 0 & \varphi_3\varphi_3^\prime \cr
& & 0
}\end{pmatrix}. \nonumber
\end{eqnarray}

Possible nestings are given below each of the linearly independent invariants. Each of them corresponds to a particular linear combination of the above invariants.

\begin{eqnarray}
(NN)_{\bf{3}}~ (\varphi\varphi^\prime)_{\bf{\bar{3}}_\pm}  &=& \m I_1^{(\varphi,\varphi^\prime)} \pm (\varphi \leftrightarrow \varphi^\prime)
\nonumber \\ 
(N\varphi)_{\bf{3}}~(N\varphi^\prime)_{\bf{\bar{3}}_\pm} &=& \m I_1^{(\varphi,\varphi^\prime)} \pm \frac{1}{2} \m I_2^{(\varphi,\varphi^\prime)}
\nonumber
\end{eqnarray}

\item There are two types of linearly independent $NN\varphi\bar{\varphi}$ invariants:

\begin{eqnarray}
\m J_1^{(\varphi,\bar{\varphi})} &=& (NN)_{\bf{3}} ~ (\varphi\bar{\varphi})_{\bf{\bar{3}}}  \longrightarrow \frac{1}{\sqrt{3}}
\begin{pmatrix}{
\varphi_2\bar{\varphi}_3 & 0 & 0 \cr
& \varphi_3\bar{\varphi}_1 & 0 \cr
& & \varphi_1\bar{\varphi}_2
}\end{pmatrix},
 \nonumber \\
 \m J_2^{(\varphi,\bar{\varphi})} &=& (NN)_{\bf{\bar{3}}} ~ (\varphi\bar{\varphi})_{\bf{3}}  \longrightarrow \frac{1}{\sqrt{6}}
\begin{pmatrix}{
0 & \varphi_1\bar{\varphi}_3 & \varphi_3\bar{\varphi}_2 \cr
& 0 & \varphi_2\bar{\varphi}_1 \cr
& & 0
}\end{pmatrix}. \nonumber
\end{eqnarray}

\begin{eqnarray}
(N\varphi)_{\bf{3}} ~ (N\bar{\varphi})_{\bf{\bar{3}}} &=& \frac{1}{\sqrt{2}} \m J_2^{(\varphi,\bar{\varphi})} 
\nonumber \\
(N\bar{\varphi})_{\bf{3}} ~ (N\varphi)_{\bf{\bar{3}}_\pm} &=& \pm \frac{1}{\sqrt{2}} \m J_1^{(\varphi,\bar{\varphi})}  + \frac{1}{2} \m J_2^{(\varphi,\bar{\varphi})} 
\nonumber
\end{eqnarray}

\item There are two types of linearly independent $NN \bar{\varphi} \bar{\varphi}^\prime$ invariants: 

\begin{eqnarray}
\m K_1^{(\bar{\varphi},\bar{\varphi}^\prime)} &=& (NN)_{\bf{3}} ~ (\bar{\varphi}\bar{\varphi}^\prime)_{\bf{\bar{3}}} \longrightarrow  \frac{1}{\sqrt{3}} \begin{pmatrix}{
\bar{\varphi}_1\bar{\varphi}_1^\prime & 0 & 0 \cr
& \bar{\varphi}_2\bar{\varphi}_2^\prime & 0 \cr
& & \bar{\varphi}_3\bar{\varphi}_3^\prime
}\end{pmatrix}, 
\nonumber \\
\m K_2^{(\bar{\varphi},\bar{\varphi}^\prime)} &=& (N\bar{\varphi})_{\bf{3}} ~ (N\bar{\varphi}^\prime)_{\bf{\bar{3}}} \longrightarrow  \frac{1}{2\sqrt{3}} \begin{pmatrix}{
0 & \bar{\varphi}_1\bar{\varphi}_2^\prime & \bar{\varphi}_3\bar{\varphi}_1^\prime \cr
& 0 & \bar{\varphi}_2\bar{\varphi}_3^\prime \cr
& & 0
}\end{pmatrix}. \nonumber 
\end{eqnarray}

\begin{eqnarray} 
(NN)_{\bf{\bar{3}}} ~ (\bar{\varphi}\bar{\varphi}^\prime)_{\bf{3_\pm}} &=& \m K_2^{(\bar{\varphi}^\prime,\bar{\varphi})} \pm (\bar{\varphi} \leftrightarrow \bar{\varphi}^\prime)
\nonumber \\
(N\bar{\varphi})_{\bf{1^\prime}} ~ (N\bar{\varphi}^\prime)_{\bf{\bar{1}^\prime}} &=& \frac{1}{\sqrt{3}} \left( \m K_1^{(\bar{\varphi},\bar{\varphi}^\prime)} + \omega \m K_2^{(\bar{\varphi},\bar{\varphi}^\prime)} + \omega^2 \m K_2^{(\bar{\varphi}^\prime,\bar{\varphi})} \right)
\nonumber \\
(N\bar{\varphi})_{\bf{1}} ~ (N\bar{\varphi}^\prime)_{\bf{1}} &=& \frac{1}{\sqrt{3}} \left( \m K_1^{(\bar{\varphi},\bar{\varphi}^\prime)} + \m K_2^{(\bar{\varphi},\bar{\varphi}^\prime)} + \m K_2^{(\bar{\varphi}^\prime,\bar{\varphi})} \right)
\nonumber
\end{eqnarray}
\end{itemize}

%%%%%%%%%%%%%%%%%%%%%%

\vskip .3cm
\noindent{\bf Dimension-six Majorana  invariants}
\vskip .3cm
\noindent We organize the possible nestings of dimension-six Majorana invariants as in the previous section, with similar notation.   

\vskip .5cm
\begin{itemize}
\item There are three types of linearly independent $NN\varphi\varphi^\prime\varphi^{\prime\prime}$ invariants:

\begin{eqnarray}
\m I_1^{(\varphi,\varphi^\prime,\varphi^{\prime\prime})} &=& \frac{1}{2} \left( \left [ (NN)_{\bf{3}}~ (\varphi\varphi^\prime)_{\bf{3}} \right ]_{\bf{\bar{3}_{+}}} \varphi^{\prime\prime} + \left [ (NN)_{\bf{3}}~ (\varphi\varphi^\prime)_{\bf{3}} \right ]_{\bf{\bar{3}_{-}}} \varphi^{\prime\prime} \right ) \nonumber \\
&\longrightarrow & \frac{1}{\sqrt{6}}
\begin{pmatrix}{
\varphi_3\varphi_3^\prime\varphi_3^{\prime\prime} & 0 & 0\cr
& \varphi_1\varphi_1^\prime\varphi_1^{\prime\prime}  & 0 \cr
& & \varphi_2\varphi_2^\prime\varphi_2^{\prime\prime} 
}\end{pmatrix}, 
\nonumber \\
\m I_2^{(\varphi,\varphi^\prime,\varphi^{\prime\prime})} &=& \frac{1}{2} \left( \left [ (NN)_{\bf{3}}~ (\varphi\varphi^\prime)_{\bf{3}} \right ]_{\bf{\bar{3}_{+}}} \varphi^{\prime\prime} - \left [ (NN)_{\bf{3}}~ (\varphi\varphi^\prime)_{\bf{3}} \right ]_{\bf{\bar{3}_{-}}} \varphi^{\prime\prime} \right ) \nonumber \\ 
&\longrightarrow & \frac{1}{\sqrt{6}}
\begin{pmatrix}{
\varphi_2\varphi_2^\prime\varphi_1^{\prime\prime} & 0 & 0\cr
&  \varphi_3\varphi_3^\prime\varphi_2^{\prime\prime} & 0 \cr
& &  \varphi_1\varphi_1^\prime\varphi_3^{\prime\prime}
}\end{pmatrix}, 
\nonumber \\
\m I_3^{(\varphi,\varphi^\prime,\varphi^{\prime\prime})} &=& \left [ (NN)_{\bf{\bar{3}}}~ (\varphi\varphi^\prime)_{\bf{3}} \right ]_{\bf{\bar{3}}} \varphi^{\prime\prime} \nonumber \\ 
&\longrightarrow  & \frac{1}{\sqrt{6}}
\begin{pmatrix}{
0 & \varphi_1\varphi_1^\prime\varphi_2^{\prime\prime} & \varphi_3\varphi_3^\prime\varphi_1^{\prime\prime} \cr
& 0 & \varphi_2\varphi_2^\prime\varphi_3^{\prime\prime} \cr
& & 0
}\end{pmatrix}. \nonumber
\end{eqnarray}

\begin{eqnarray}
%-------------------------------------------------------------
%(NN)(\varphi\varphi^\prime) \varphi^{\prime\prime}
%-------------------------------------------------------------
\left [ (NN)_{\bf{3}}~ (\varphi\varphi^\prime)_{\bf{3}} \right ]_{\bf{\bar{3}_{\pm}}} \varphi^{\prime\prime} &=& \m I_1^{(\varphi,\varphi^\prime, \varphi^{\prime\prime})} \pm \m I_2^{(\varphi,\varphi^\prime, \varphi^{\prime\prime})} 
\nonumber \\
\left [ (NN)_{\bf{3}}~ (\varphi\varphi^\prime)_{\bf{\bar{3}_\pm}} \right ]_{\bf{\bar{3}}} \varphi^{\prime\prime} &=& \m I_2^{(\varphi,\varphi^{\prime\prime}, \varphi^{\prime})} \pm \m I_2^{(\varphi^{\prime\prime},\varphi^{\prime}, \varphi^{})} 
\nonumber \\
\left [ (NN)_{\bf{\bar{3}}}~ (\varphi\varphi^\prime)_{\bf{\bar{3}_\pm}} \right ]_{\bf{\bar{3}}} \varphi^{\prime\prime} &=& \frac{1}{\sqrt{2}} \left( \m I_3^{(\varphi^{\prime\prime},\varphi^{\prime}, \varphi^{})} \pm \m I_3^{(\varphi^{},\varphi^{\prime\prime}, \varphi^{\prime})} \right)
\nonumber \\
%-------------------------------------------------------------
%(N\varphi)(\varphi^\prime\varphi^{\prime\prime}) N
%-------------------------------------------------------------
\left [ (N\varphi)_{\bf{3}}~ (\varphi^\prime\varphi^{\prime\prime})_{\bf{3}} \right ]_{\bf{\bar{3}_\pm}} N &=& \pm \m I_2^{(\varphi^{\prime\prime},\varphi^{\prime}, \varphi^{})} + \frac{1}{2} \m I_3^{(\varphi^{\prime\prime},\varphi^{\prime}, \varphi^{})} 
\nonumber\\
\left [ (N\varphi^{})_{\bf{3}}~ (\varphi^{\prime}\varphi^{\prime\prime})_{\bf{\bar{3}_\pm}} \right ]_{\bf{\bar{3}}} N &=& \frac{1}{2} \m I_3^{(\varphi^{},\varphi^{\prime\prime}, \varphi^{\prime})} \pm (\varphi^\prime \leftrightarrow \varphi^{\prime\prime})  
\nonumber \\
\left [ (N\varphi)_{\bf{\bar{3}_\pm}}~ (\varphi^\prime\varphi^{\prime\prime})_{\bf{3}} \right ]_{\bf{\bar{3}}} N &=& \m I_1^{(\varphi^{},\varphi^{\prime}, \varphi^{\prime\prime})} \pm \frac{1}{2} \m I_3^{(\varphi^{\prime\prime},\varphi^{\prime}, \varphi^{})} 
\nonumber \\
\left [ (N\varphi^{})_{\bf{\bar{3}_\pm}}~ (\varphi^\prime\varphi^{\prime\prime})_{\bf{\bar{3}_+}} \right ]_{\bf{\bar{3}}} N &=& \frac{1}{\sqrt{2}} \left( \pm \m I_2^{(\varphi^{},\varphi^{\prime}, \varphi^{\prime\prime})} + \frac{1}{2}  \m I_3^{(\varphi^{},\varphi^{\prime\prime}, \varphi^{\prime})} \right) + (\varphi^\prime \leftrightarrow \varphi^{\prime\prime})  
\nonumber\\
\left [ (N\varphi^{})_{\bf{\bar{3}_\pm}}~ (\varphi^\prime\varphi^{\prime\prime})_{\bf{\bar{3}_-}} \right ]_{\bf{\bar{3}}} N &=& \frac{1}{\sqrt{2}} \left( \pm \m I_2^{(\varphi^{},\varphi^{\prime}, \varphi^{\prime\prime})} + \frac{1}{2}  \m I_3^{(\varphi^{},\varphi^{\prime\prime}, \varphi^{\prime})} \right) - (\varphi^\prime \leftrightarrow \varphi^{\prime\prime})   
\nonumber\\
%-------------------------------------------------------------
%(N\varphi)(N\varphi^\prime) \varphi^{\prime\prime}
%-------------------------------------------------------------
\left [ (N\varphi)_{\bf{3}}~ (N\varphi^\prime)_{\bf{3}} \right ]_{\bf{\bar{3}_\pm}} \varphi^{\prime\prime} &=& \frac{1}{2} \left( \m I_3^{(\varphi^{\prime\prime},\varphi^{\prime}, \varphi^{})} \pm \m I_3^{(\varphi^{},\varphi^{\prime\prime}, \varphi^{\prime})}  \right) \nonumber \\
\left [ (N\varphi)_{\bf{3}}~ (N\varphi^\prime)_{\bf{\bar{3}_\pm}} \right ]_{\bf{\bar{3}}} \varphi^{\prime\prime} &=& \pm \m I_2^{(\varphi^{\prime\prime},\varphi^{\prime}, \varphi^{})} + \frac{1}{2} \m I_3^{(\varphi^{},\varphi^{\prime}, \varphi^{\prime\prime})} \nonumber \\
\left [ (N\varphi)_{\bf{\bar{3}_+}}~ (N\varphi^\prime)_{\bf{\bar{3}_\pm}} \right ]_{\bf{\bar{3}}} \varphi^{\prime\prime} &=& \frac{1}{\sqrt{2}} \left(\pm \m I_2^{(\varphi^{},\varphi^{\prime}, \varphi^{\prime\prime})} + \m I_1^{(\varphi^{},\varphi^{\prime}, \varphi^{\prime\prime})} \right) + \frac{1}{2\sqrt{2}} \left(\m I_3^{(\varphi^{\prime\prime},\varphi^{\prime}, \varphi^{})} \pm \m I_3^{(\varphi^{},\varphi^{\prime\prime}, \varphi^{\prime})}  \right)\nonumber \\
\left [ (N\varphi)_{\bf{\bar{3}_-}}~ (N\varphi^\prime)_{\bf{\bar{3}_\pm}} \right ]_{\bf{\bar{3}}} \varphi^{\prime\prime} &=& \frac{1}{\sqrt{2}} \left(\mp \m I_2^{(\varphi^{},\varphi^{\prime}, \varphi^{\prime\prime})} + \m I_1^{(\varphi^{},\varphi^{\prime}, \varphi^{\prime\prime})} \right) + \frac{1}{2\sqrt{2}} \left(-\m I_3^{(\varphi^{\prime\prime},\varphi^{\prime}, \varphi^{})} \pm \m I_3^{(\varphi^{},\varphi^{\prime\prime}, \varphi^{\prime})}  \right) \nonumber 
\end{eqnarray}

\vskip .5cm
\item There are five types of linearly independent $NN\varphi\varphi^\prime\bar{\varphi}$ invariants:

\begin{eqnarray}
\m J_1^{(\varphi,\bar{\varphi},\varphi^\prime)} &=& \left [ (NN)_{\bf{3}}~ (\bar{\varphi}\varphi)_{\bf{\bar{3}}} \right ]_{\bf{\bar{3}}} \varphi^\prime \longrightarrow \frac{1}{\sqrt{3}}
\begin{pmatrix}{
\varphi_3\bar{\varphi}_1\varphi_2^\prime & 0 & 0 \cr
& \varphi_1\bar{\varphi}_2\varphi_3^\prime & 0 \cr
& & \varphi_2\bar{\varphi}_3\varphi_1^\prime
}\end{pmatrix},
 \nonumber \\
\m J_2^{(\varphi,\bar{\varphi},\varphi^\prime)} &=& \left [ (NN)_{\bf{3}}~ (\varphi\varphi^\prime)_{\bf{3}} \right ]_{\bf{3}} \bar{\varphi} \longrightarrow \frac{1}{\sqrt{3}}
\begin{pmatrix}{
\varphi_1\bar{\varphi}_3\varphi_1^\prime & 0 & 0 \cr
& \varphi_2\bar{\varphi}_1\varphi_2^\prime & \cr
& & \varphi_3\bar{\varphi}_2\varphi_3^\prime
}\end{pmatrix},
 \nonumber \\
\m J_3^{(\varphi,\bar{\varphi},\varphi^\prime)} &=& \left [ (NN)_{\bf{\bar{3}}}~ (\bar{\varphi}\varphi)_{\bf{3}} \right ]_{\bf{\bar{3}}} \varphi^\prime \longrightarrow \frac{1}{\sqrt{6}}
\begin{pmatrix}{
0 & \varphi_3\bar{\varphi}_2\varphi_2^\prime & \varphi_2\bar{\varphi}_1\varphi_1^\prime \cr
& 0 & \varphi_1\bar{\varphi}_3\varphi_3^\prime \cr
& & 0
}\end{pmatrix},
 \nonumber \\
\m J_4^{(\varphi,\bar{\varphi},\varphi^\prime)} &=& \left [ (NN)_{\bf{\bar{3}}}~ (\bar{\varphi}\varphi)_{\bf{\bar{3}}} \right ]_{\bf{\bar{3}}} \varphi^\prime \longrightarrow \frac{1}{\sqrt{6}}
\begin{pmatrix}{
0 & \varphi_3\bar{\varphi}_1\varphi_1^\prime & \varphi_2\bar{\varphi}_3\varphi_3^\prime \cr
& 0 & \varphi_1\bar{\varphi}_2\varphi_2^\prime \cr
& & 0
}\end{pmatrix},
 \nonumber \\
\m J_5^{(\varphi,\bar{\varphi},\varphi^\prime)} &=& \left [ (NN)_{\bf{\bar{3}}}~ (\varphi\varphi^\prime)_{\bf{3}} \right ]_{\bf{3}} \bar{\varphi} \longrightarrow \frac{1}{\sqrt{6}}
\begin{pmatrix}{
0 & \varphi_3\bar{\varphi}_3\varphi_3^\prime & \varphi_2\bar{\varphi}_2\varphi_2^\prime \cr
& 0 & \varphi_1\bar{\varphi}_1\varphi_1^\prime \cr
& & 0
}\end{pmatrix}. \nonumber
\end{eqnarray}

\begin{eqnarray}
%-----------------------------
%(NN)(\bar{\varphi}\varphi) \varphi^\prime
\left [ (NN)_{\bf{3}}~ (\bar{\varphi}\varphi)_{\bf{3}} \right ]_{\bf{\bar{3}_{\pm}}} \varphi^\prime &=& \frac{1}{\sqrt{2}} \left( \pm \m J_2^{(\varphi^{}, \bar{\varphi}, \varphi^{\prime})} + \m J_1^{(\varphi^{\prime}, \bar{\varphi}, \varphi^{})} \right) 
\nonumber \\
%-------------------------------------------------------------
%(NN)(\varphi\varphi^\prime) \bar{\varphi}
%-------------------------------------------------------------
\left [ (NN)_{\bf{3}}~ (\varphi\varphi^\prime)_{\bf{\bar{3}_\pm}} \right ]_{\bf{3}} \bar{\varphi} &=& \frac{1}{\sqrt{2}} \m J_1^{(\varphi^{}, \bar{\varphi}, \varphi^{\prime})} \pm (\varphi \leftrightarrow \varphi^\prime)  
\nonumber \\
\left [ (NN)_{\bf{\bar{3}}}~ (\varphi\varphi^\prime)_{\bf{\bar{3}_+}} \right ]_{\bf{3_\pm}} \bar{\varphi} &=& \frac{1}{2} \left( \m J_4^{(\varphi^{}, \bar{\varphi}, \varphi^{\prime})} \pm \m J_3^{(\varphi^{}, \bar{\varphi}, \varphi^{\prime})} \right) + ( \varphi \leftrightarrow \varphi^\prime )  
\nonumber  \\
\left [ (NN)_{\bf{\bar{3}}}~ (\varphi\varphi^\prime)_{\bf{\bar{3}_-}} \right ]_{\bf{3_\pm}} \bar{\varphi} &=& \frac{1}{2} \left( - \m J_4^{(\varphi^{}, \bar{\varphi}, \varphi^{\prime})} \pm \m J_3^{(\varphi^{}, \bar{\varphi}, \varphi^{\prime})} \right) - ( \varphi \leftrightarrow \varphi^\prime )  
 \nonumber  \\
%-------------------------------------------------------------
%(N\bar{\varphi})(N\varphi) \varphi^\prime
%-------------------------------------------------------------
\left [ (N\bar{\varphi})_{\bf{3}}~ (N\varphi)_{\bf{3}} \right ]_{\bf{\bar{3}_\pm}} \varphi^\prime &=& \frac{1}{\sqrt{2}} \m J_2^{(\varphi^{}, \bar{\varphi}, \varphi^{\prime})} \pm \frac{1}{2} \m J_4^{(\varphi^{\prime}, \bar{\varphi}, \varphi^{})} \nonumber \\
\left [ (N\bar{\varphi})_{\bf{3}}~ (N\varphi)_{\bf{\bar{3}_\pm}} \right ]_{\bf{\bar{3}}} \varphi^\prime &=& \frac{1}{2} \left( \m J_4^{(\varphi^{}, \bar{\varphi}, \varphi^{\prime})} \pm \m J_5^{(\varphi^{}, \bar{\varphi}, \varphi^{\prime})} \right) \nonumber \\
\left [ (N\bar{\varphi})_{\bf{\bar{3}}}~ (N\varphi)_{\bf{3}} \right ]_{\bf{\bar{3}}} \varphi^\prime &=& \frac{1}{\sqrt{2}} \m J_3^{(\varphi^{\prime}, \bar{\varphi}, \varphi^{})} \nonumber \\
\left [ (N\bar{\varphi})_{\bf{\bar{3}}}~ (N\varphi)_{\bf{\bar{3}_\pm}} \right ]_{\bf{\bar{3}}} \varphi^\prime &=& \frac{1}{2} \left( \pm \m J_3^{(\varphi^{}, \bar{\varphi}, \varphi^{\prime})} + \m J_5^{(\varphi^{}, \bar{\varphi}, \varphi^{\prime})} \right) \nonumber \\
%-------------------------------------------------------------
%(N\bar{\varphi})(\varphi\varphi^\prime) N
%-------------------------------------------------------------
\left [ (N\bar{\varphi})_{\bf{3}}~ (\varphi\varphi^\prime)_{\bf{3}} \right ]_{\bf{\bar{3}_\pm}} N &=& \frac{1}{\sqrt{2}} \m J_2^{(\varphi^{}, \bar{\varphi}, \varphi^{\prime})} \pm \frac{1}{2} \m J_5^{(\varphi^{}, \bar{\varphi}, \varphi^{\prime})} 
\nonumber \\
\left [ (N\bar{\varphi})_{\bf{3}}~ (\varphi\varphi^\prime)_{\bf{\bar{3}_\pm}} \right ]_{\bf{\bar{3}}} N &=& \frac{1}{2} \m J_4^{(\varphi^{\prime}, \bar{\varphi}, \varphi^{})}  \pm (\varphi \leftrightarrow \varphi^\prime)
\nonumber \\
\left [ (N\bar{\varphi})_{\bf{\bar{3}}}~ (\varphi\varphi^\prime)_{\bf{3}} \right ]_{\bf{\bar{3}}} N &=& \frac{1}{\sqrt{2}} \m J_5^{(\varphi^{}, \bar{\varphi}, \varphi^{\prime})} 
\nonumber \\
\left [ (N\bar{\varphi})_{\bf{\bar{3}}}~ (\varphi\varphi^\prime)_{\bf{\bar{3}_\pm}} \right ]_{\bf{\bar{3}}} N &=& \frac{1}{2} \m J_3^{(\varphi^{}, \bar{\varphi}, \varphi^{\prime})} \pm (\varphi \leftrightarrow \varphi^\prime) \nonumber \\
%-------------------------------------------------------------
%(N\varphi)(\bar{\varphi}\varphi^\prime) N
%-------------------------------------------------------------
\left [ (N\varphi)_{\bf{3}}~ (\bar{\varphi}\varphi^\prime)_{\bf{3}} \right ]_{\bf{\bar{3}_\pm}} N &=& \pm \frac{1}{\sqrt{2}} \m J_2^{(\varphi^{}, \bar{\varphi}, \varphi^{\prime})} + \frac{1}{2} \m J_3^{(\varphi^{\prime}, \bar{\varphi}, \varphi^{})}  \nonumber \\
\left [ (N\varphi)_{\bf{3}}~ (\bar{\varphi}\varphi^\prime)_{\bf{\bar{3}}} \right ]_{\bf{\bar{3}}} N &=& \frac{1}{\sqrt{2}} \m J_4^{(\varphi^{\prime}, \bar{\varphi}, \varphi^{})} \nonumber \\
\left [ (N\varphi)_{\bf{\bar{3}_\pm}}~ (\bar{\varphi}\varphi^\prime)_{\bf{3}} \right ]_{\bf{\bar{3}}} N &=&\frac{1}{\sqrt{2}} \m J_1^{(\varphi^{}, \bar{\varphi}, \varphi^{\prime})} \pm \frac{1}{2} \m J_3^{(\varphi^{\prime}, \bar{\varphi}, \varphi^{})} \nonumber \\
\left [ (N\varphi)_{\bf{\bar{3}_\pm}}~ (\bar{\varphi}\varphi^\prime)_{\bf{\bar{3}}} \right ]_{\bf{\bar{3}}} N &=& \pm \frac{1}{\sqrt{2}} \m J_1^{(\varphi^{\prime}, \bar{\varphi}, \varphi^{})} + \frac{1}{2} \m J_4^{(\varphi^{\prime}, \bar{\varphi}, \varphi^{})}\nonumber \\
%-------------------------------------------------------------
%(N\varphi)(N\varphi^\prime) \bar{\varphi}
%-------------------------------------------------------------
\left [ (N\varphi)_{\bf{3}}~ (N\varphi^\prime)_{\bf{3}} \right ]_{\bf{3}} \bar{\varphi} &=& \m J_2^{(\varphi^{}, \bar{\varphi}, \varphi^{\prime})} \nonumber \\
\left [ (N\varphi)_{\bf{3}}~ (N\varphi^\prime)_{\bf{\bar{3}_\pm}} \right ]_{\bf{3}} \bar{\varphi} &=& \frac{1}{2} \left( \pm \m J_4^{(\varphi^{\prime}, \bar{\varphi}, \varphi^{})} + \m J_3^{(\varphi^{\prime}, \bar{\varphi}, \varphi^{})} \right) 
\nonumber \\
\left [ (N\varphi)_{\bf{\bar{3}_+}}~ (N\varphi^\prime)_{\bf{\bar{3}_+}} \right ]_{\bf{3_+}} \bar{\varphi} &=& \frac{1}{2\sqrt{2}} \left( \m J_1^{(\varphi^{}, \bar{\varphi}, \varphi^{\prime})} + \m J_1^{(\varphi^{\prime}, \bar{\varphi}, \varphi^{})} \right) + 
\nonumber \\
& &  \frac{1}{4} \left( \m J_4^{(\varphi^{\prime}, \bar{\varphi}, \varphi^{})} + \m J_3^{(\varphi^{\prime}, \bar{\varphi}, \varphi^{})} + \m J_4^{(\varphi^{}, \bar{\varphi}, \varphi^{\prime})} + \m J_3^{(\varphi^{}, \bar{\varphi}, \varphi^{\prime})} + 2 \m J_5^{(\varphi^{}, \bar{\varphi}, \varphi^{\prime})} \right) 
\nonumber \\
\left [ (N\varphi)_{\bf{\bar{3}_+}}~ (N\varphi^\prime)_{\bf{\bar{3}_-}} \right ]_{\bf{3_+}} \bar{\varphi} &=& \frac{1}{2\sqrt{2}} \left( - \m J_1^{(\varphi^{}, \bar{\varphi}, \varphi^{\prime})} + \m J_1^{(\varphi^{\prime}, \bar{\varphi}, \varphi^{})} \right) + \nonumber \\
& &   \frac{1}{4} \left( \m J_4^{(\varphi^{\prime}, \bar{\varphi}, \varphi^{})} - \m J_3^{(\varphi^{\prime}, \bar{\varphi}, \varphi^{})} + \m J_4^{(\varphi^{}, \bar{\varphi}, \varphi^{\prime})} - \m J_3^{(\varphi^{}, \bar{\varphi}, \varphi^{\prime})}  \right) 
\nonumber \\
\left [ (N\varphi)_{\bf{\bar{3}_-}}~ (N\varphi^\prime)_{\bf{\bar{3}_+}} \right ]_{\bf{3_+}} \bar{\varphi} &=& \frac{1}{2\sqrt{2}} \left(  \m J_1^{(\varphi^{}, \bar{\varphi}, \varphi^{\prime})} - \m J_1^{(\varphi^{\prime}, \bar{\varphi}, \varphi^{})} \right)+ \nonumber \\
& &  \frac{1}{4} \left( \m J_4^{(\varphi^{\prime}, \bar{\varphi}, \varphi^{})} - \m J_3^{(\varphi^{\prime}, \bar{\varphi}, \varphi^{})} + \m J_4^{(\varphi^{}, \bar{\varphi}, \varphi^{\prime})} - \m J_3^{(\varphi^{}, \bar{\varphi}, \varphi^{\prime})}  \right) 
\nonumber \\
\left [ (N\varphi)_{\bf{\bar{3}_-}}~ (N\varphi^\prime)_{\bf{\bar{3}_-}} \right ]_{\bf{3_+}} \bar{\varphi} &=& \frac{1}{2\sqrt{2}} \left( -\m J_1^{(\varphi^{}, \bar{\varphi}, \varphi^{\prime})} - \m J_1^{(\varphi^{\prime}, \bar{\varphi}, \varphi^{})} \right) +\nonumber \\
& & \frac{1}{4} \left( \m J_4^{(\varphi^{\prime}, \bar{\varphi}, \varphi^{})} + \m J_3^{(\varphi^{\prime}, \bar{\varphi}, \varphi^{})} + \m J_4^{(\varphi^{}, \bar{\varphi}, \varphi^{\prime})} + \m J_3^{(\varphi^{}, \bar{\varphi}, \varphi^{\prime})} - 2 \m J_5^{(\varphi^{}, \bar{\varphi}, \varphi^{\prime})} \right) 
\nonumber \\
\left [ (N\varphi)_{\bf{\bar{3}_+}}~ (N\varphi^\prime)_{\bf{\bar{3}_+}} \right ]_{\bf{3_-}} \bar{\varphi} &=& \frac{1}{2\sqrt{2}} \left( -\m J_1^{(\varphi^{}, \bar{\varphi}, \varphi^{\prime})} + \m J_1^{(\varphi^{\prime}, \bar{\varphi}, \varphi^{})} \right) +\nonumber \\
& &  \frac{1}{4} \left( \m J_4^{(\varphi^{\prime}, \bar{\varphi}, \varphi^{})} - \m J_3^{(\varphi^{\prime}, \bar{\varphi}, \varphi^{})} - \m J_4^{(\varphi^{}, \bar{\varphi}, \varphi^{\prime})} + \m J_3^{(\varphi^{}, \bar{\varphi}, \varphi^{\prime})}  \right) 
\nonumber \\
\left [ (N\varphi)_{\bf{\bar{3}_+}}~ (N\varphi^\prime)_{\bf{\bar{3}_-}} \right ]_{\bf{3_-}} \bar{\varphi} &=& \frac{1}{2\sqrt{2}} \left( \m J_1^{(\varphi^{}, \bar{\varphi}, \varphi^{\prime})} + \m J_1^{(\varphi^{\prime}, \bar{\varphi}, \varphi^{})} \right)+ \nonumber \\
& &   \frac{1}{4} \left( \m J_4^{(\varphi^{\prime}, \bar{\varphi}, \varphi^{})} + \m J_3^{(\varphi^{\prime}, \bar{\varphi}, \varphi^{})} - \m J_4^{(\varphi^{}, \bar{\varphi}, \varphi^{\prime})} - \m J_3^{(\varphi^{}, \bar{\varphi}, \varphi^{\prime})} - 2 \m J_5^{(\varphi^{}, \bar{\varphi}, \varphi^{\prime})} \right) \nonumber \\
\left [ (N\varphi)_{\bf{\bar{3}_-}}~ (N\varphi^\prime)_{\bf{\bar{3}_+}} \right ]_{\bf{3_-}} \bar{\varphi} &=& \frac{1}{2\sqrt{2}} \left( -\m J_1^{(\varphi^{}, \bar{\varphi}, \varphi^{\prime})} - \m J_1^{(\varphi^{\prime}, \bar{\varphi}, \varphi^{})} \right)+ \nonumber \\
& &  \frac{1}{4} \left( \m J_4^{(\varphi^{\prime}, \bar{\varphi}, \varphi^{})} + \m J_3^{(\varphi^{\prime}, \bar{\varphi}, \varphi^{})} - \m J_4^{(\varphi^{}, \bar{\varphi}, \varphi^{\prime})} - \m J_3^{(\varphi^{}, \bar{\varphi}, \varphi^{\prime})} + 2 \m J_5^{(\varphi^{}, \bar{\varphi}, \varphi^{\prime})} \right) 
\nonumber \\
\left [ (N\varphi)_{\bf{\bar{3}_-}}~ (N\varphi^\prime)_{\bf{\bar{3}_-}} \right ]_{\bf{3_-}} \bar{\varphi} &=& \frac{1}{2\sqrt{2}} \left( \m J_1^{(\varphi^{}, \bar{\varphi}, \varphi^{\prime})} - \m J_1^{(\varphi^{\prime}, \bar{\varphi}, \varphi^{})} \right) +\nonumber \\
& &  \frac{1}{4} \left( \m J_4^{(\varphi^{\prime}, \bar{\varphi}, \varphi^{})} - \m J_3^{(\varphi^{\prime}, \bar{\varphi}, \varphi^{})} - \m J_4^{(\varphi^{}, \bar{\varphi}, \varphi^{\prime})} + \m J_3^{(\varphi^{}, \bar{\varphi}, \varphi^{\prime})} \right) 
 \nonumber 
\end{eqnarray}
\vskip .5cm
\item There are five types of linearly independent $NN \bar{\varphi} \bar{\varphi}^\prime \varphi$ invariants: 

\begin{eqnarray}
\m K_1^{(\varphi,\bar{\varphi},\bar{\varphi}^\prime)} &=& \left[ (NN)_{\bf{3}} (\bar{\varphi} \varphi)_{\bf{3}} \right]_{\bf{3}} \bar{\varphi^\prime}  \longrightarrow  \frac{1}{\sqrt{3}}\begin{pmatrix}{ \varphi_3 \bar{\varphi}_2 \bar{\varphi}_3^{\prime} & 0 & 0 \cr & \varphi_1 \bar{\varphi}_3 \bar{\varphi}_1^{\prime} & 0 \cr & & \varphi_2 \bar{\varphi}_1 \bar{\varphi}_2^{\prime}
}\end{pmatrix},  \nonumber \\
\m K_2^{(\varphi,\bar{\varphi},\bar{\varphi}^\prime)} &=& \left[ (NN)_{\bf{3}} (\bar{\varphi} \varphi)_{\bf{\bar{3}}} \right]_{\bf{3}} \bar{\varphi^\prime}  \longrightarrow \frac{1}{\sqrt{3}} \begin{pmatrix}{ \varphi_1 \bar{\varphi}_2 \bar{\varphi}_1^{\prime} & 0 & 0 \cr & \varphi_2 \bar{\varphi}_3 \bar{\varphi}_2^{\prime} & 0 \cr & & \varphi_3 \bar{\varphi}_1 \bar{\varphi}_3^{\prime}
}\end{pmatrix}, \nonumber \\
\m K_3^{(\varphi,\bar{\varphi},\bar{\varphi}^\prime)} &=& \left[ (NN)_{\bf{3}} (\bar{\varphi} \bar{\varphi}^\prime)_{\bf{\bar{3}}} \right]_{\bf{\bar{3}}} \varphi  \longrightarrow \frac{1}{\sqrt{3}}  \begin{pmatrix}{ \varphi_2 \bar{\varphi}_2 \bar{\varphi}_2^{\prime} & 0 & 0 \cr & \varphi_3 \bar{\varphi}_3 \bar{\varphi}_3^{\prime} & 0 \cr & & \varphi_1 \bar{\varphi}_1 \bar{\varphi}_1^{\prime}
}\end{pmatrix},  \nonumber \\
\m K_4^{(\varphi,\bar{\varphi},\bar{\varphi}^\prime)} &=& \left[ (NN)_{\bf{\bar{3}}} (\bar{\varphi} \varphi)_{\bf{3}} \right]_{\bf{3}} \bar{\varphi^\prime} \longrightarrow \frac{1}{\sqrt{6}} \begin{pmatrix}{ 
0 & \varphi_2\bar{\varphi}_1\bar{\varphi}_3^\prime & \varphi_1\bar{\varphi}_3\bar{\varphi}_2^\prime \cr
& 0 & \varphi_3\bar{\varphi}_2\bar{\varphi}_1^\prime \cr
& & 0
}\end{pmatrix}, \nonumber \\
\m K_5^{(\varphi,\bar{\varphi},\bar{\varphi}^\prime)} &=& \left[ (NN)_{\bf{\bar{3}}} (\bar{\varphi} \bar{\varphi}^\prime)_{\bf{\bar{3}}} \right]_{\bf{\bar{3}}} \varphi  \longrightarrow  \frac{1}{\sqrt{6}} \begin{pmatrix}{ 
0 & \varphi_1\bar{\varphi}_2\bar{\varphi}_2^\prime & \varphi_3\bar{\varphi}_1\bar{\varphi}_1^\prime \cr
& 0 & \varphi_2\bar{\varphi}_3\bar{\varphi}_3^\prime \cr
& & 0
}\end{pmatrix}.  \nonumber
\end{eqnarray}

\begin{eqnarray} 
%-------------------------------------------
%%%%%%%% (NN) (\bar{\varphi}\varphi) \bar{\varphi}^\prime
%---------------------------------------------
\left[ (NN)_{\bf{\bar{3}}} (\bar{\varphi} \varphi)_{\bf{\bar{3}}} \right]_{\bf{3_\pm}} \bar{\varphi^\prime} &=& \frac{1}{\sqrt{2}} \left( \m K_4^{(\varphi, \bar{\varphi}^{\prime},\bar{\varphi}^{})} \pm \m K_5^{(\varphi, \bar{\varphi}^{},\bar{\varphi}^{\prime})} \right)   \nonumber \\
%------------------------------------------------
%%%%%%%% (NN) (\bar{\varphi}\bar{\varphi}^\prime) \varphi
%----------------------------------------------------
\left[ (NN)_{\bf{3}} (\bar{\varphi} \bar{\varphi}^\prime)_{\bf{3_+}} \right]_{\bf{\bar{3}_\pm}} \varphi &=& \frac{1}{2} \left( \pm \m K_2^{(\varphi, \bar{\varphi}^{},\bar{\varphi}^{\prime})} +  \m K_1^{(\varphi, \bar{\varphi}^{\prime},\bar{\varphi}^{})} \right) + (\bar{\varphi} \leftrightarrow \bar{\varphi}^\prime)
\nonumber \\
\left[ (NN)_{\bf{3}} (\bar{\varphi} \bar{\varphi}^\prime)_{\bf{3_-}} \right]_{\bf{\bar{3}_\pm}} \varphi &=& \frac{1}{2} \left( \pm \m K_2^{(\varphi, \bar{\varphi}^{},\bar{\varphi}^{\prime})} +  \m K_1^{(\varphi, \bar{\varphi}^{\prime},\bar{\varphi}^{})} \right) - (\bar{\varphi} \leftrightarrow \bar{\varphi}^\prime)
\nonumber \\
\left[ (NN)_{\bf{\bar{3}}} (\bar{\varphi} \bar{\varphi}^\prime)_{\bf{3_\pm}} \right]_{\bf{\bar{3}}} \varphi &=& \frac{1}{\sqrt{2}} \m K_4^{(\varphi, \bar{\varphi}^{},\bar{\varphi}^{\prime})} \pm  (\bar{\varphi} \leftrightarrow \bar{\varphi}^\prime) \nonumber \\
%-------------------------------------------------------
%%%%%%%%%   (N \bar{\varphi}) (N \bar{\varphi}) \varphi
%-------------------------------------------------------
\left[ (N \bar{\varphi})_{\bf{3}} (N \bar{\varphi}^\prime)_{\bf{3}} \right]_{\bf{\bar{3}_\pm}} \varphi &=& \frac{1}{2} \m K_4^{(\varphi, \bar{\varphi}^{},\bar{\varphi}^{\prime})} \pm  (\bar{\varphi} \leftrightarrow \bar{\varphi}^\prime)
\nonumber \\ 
\left[ (N \bar{\varphi})_{\bf{3}} (N \bar{\varphi}^\prime)_{\bf{\bar{3}}} \right]_{\bf{\bar{3}}} \varphi &=& \m K_1^{(\varphi, \bar{\varphi}^{\prime},\bar{\varphi}^{})} 
\nonumber \\
\left[ (N \bar{\varphi})_{\bf{\bar{3}}} (N \bar{\varphi}^\prime)_{\bf{\bar{3}}} \right]_{\bf{\bar{3}}} \varphi &=& \m K_3^{(\varphi, \bar{\varphi}^{},\bar{\varphi}^{\prime})}  \nonumber \\
%%-------------------------------------------------------
%%% (N\bar{\varphi}) (N\varphi) \bar{\varphi}^\prime
%%---------------------------------------------------
\left[ (N \bar{\varphi})_{\bf{3}} (N \varphi)_{\bf{3}} \right]_{\bf{3}} \bar{\varphi}^\prime &=& \frac{1}{\sqrt{2}} \m K_4^{(\varphi, \bar{\varphi}^{\prime},\bar{\varphi}^{})} 
\nonumber \\
\left[ (N \bar{\varphi})_{\bf{3}} (N \varphi)_{\bf{\bar{3}}_\pm} \right]_{\bf{3}} \bar{\varphi}^\prime &=& \frac{1}{\sqrt{2}} \m K_1^{(\varphi, \bar{\varphi}^{\prime},\bar{\varphi}^{})} \pm \frac{1}{2} \m K_4^{(\varphi, \bar{\varphi}^{},\bar{\varphi}^{\prime})} 
\nonumber \\
\left[ (N \bar{\varphi})_{\bf{\bar{3}}} (N \varphi)_{\bf{3}} \right]_{\bf{3}} \bar{\varphi}^\prime &=& \m K_2^{(\varphi, \bar{\varphi}^{},\bar{\varphi}^{\prime})} 
\nonumber \\
\left[ (N \bar{\varphi})_{\bf{\bar{3}}} (N \varphi)_{\bf{\bar{3}}_+} \right]_{\bf{3}_+} \bar{\varphi}^\prime &=& \frac{1}{2} \left( \m K_3^{(\varphi, \bar{\varphi}^{},\bar{\varphi}^{\prime})} + \m K_1^{(\varphi, \bar{\varphi}^{},\bar{\varphi}^{\prime})} \right) + \frac{1}{2\sqrt{2}} \left( \m K_4^{(\varphi, \bar{\varphi}^{\prime},\bar{\varphi}^{})} + \m K_5^{(\varphi, \bar{\varphi}^{},\bar{\varphi}^{\prime})} \right) 
\nonumber \\
\left[ (N \bar{\varphi})_{\bf{\bar{3}}} (N \varphi)_{\bf{\bar{3}}_+} \right]_{\bf{3}_-} \bar{\varphi}^\prime &=& \frac{1}{2} \left( \m K_3^{(\varphi, \bar{\varphi}^{},\bar{\varphi}^{\prime})} - \m K_1^{(\varphi, \bar{\varphi}^{},\bar{\varphi}^{\prime})} \right) + \frac{1}{2\sqrt{2}} \left( - \m K_4^{(\varphi, \bar{\varphi}^{\prime},\bar{\varphi}^{})} + \m K_5^{(\varphi, \bar{\varphi}^{},\bar{\varphi}^{\prime})} \right)
 \nonumber \\
\left[ (N \bar{\varphi})_{\bf{\bar{3}}} (N \varphi)_{\bf{\bar{3}}_-} \right]_{\bf{3}_+} \bar{\varphi}^\prime &=& \frac{1}{2} \left(- \m K_3^{(\varphi, \bar{\varphi}^{},\bar{\varphi}^{\prime})} + \m K_1^{(\varphi, \bar{\varphi}^{},\bar{\varphi}^{\prime})} \right) + \frac{1}{2\sqrt{2}} \left( - \m K_4^{(\varphi, \bar{\varphi}^{\prime},\bar{\varphi}^{})} + \m K_5^{(\varphi, \bar{\varphi}^{},\bar{\varphi}^{\prime})} \right)
\nonumber \\
\left[ (N \bar{\varphi})_{\bf{\bar{3}}} (N \varphi)_{\bf{\bar{3}}_-} \right]_{\bf{3}_-} \bar{\varphi}^\prime &=& \frac{1}{2} \left( -\m K_3^{(\varphi, \bar{\varphi}^{},\bar{\varphi}^{\prime})} - \m K_1^{(\varphi, \bar{\varphi}^{},\bar{\varphi}^{\prime})} \right) + \frac{1}{2\sqrt{2}} \left( \m K_4^{(\varphi, \bar{\varphi}^{\prime},\bar{\varphi}^{})} + \m K_5^{(\varphi, \bar{\varphi}^{},\bar{\varphi}^{\prime})} \right)
\nonumber \\
%%-----------------------------------------------------------
%%%%%%%%%%%  (N \bar{\varphi}) (\varphi \bar{\varphi}) N
%%-----------------------------------------------------------
\left[ (N \bar{\varphi})_{\bf{3}} (\varphi \bar{\varphi}^\prime)_{\bf{3}} \right]_{\bf{\bar{3}_\pm}} N &=& \frac{1}{\sqrt{2}}  \m K_1^{(\varphi, \bar{\varphi}^{\prime},\bar{\varphi}^{})} \pm \frac{1}{2} \m K_4^{(\varphi, \bar{\varphi}^{\prime},\bar{\varphi}^{})} \nonumber \\
\left[ (N \bar{\varphi})_{\bf{3}} (\varphi \bar{\varphi}^\prime)_{\bf{\bar{3}}} \right]_{\bf{\bar{3}}} N &=& \frac{1}{\sqrt{2}} \m K_4^{(\varphi, \bar{\varphi}^{},\bar{\varphi}^{\prime})}  
\nonumber \\
\left[ (N \bar{\varphi})_{\bf{\bar{3}}} (\varphi \bar{\varphi}^\prime)_{\bf{3}} \right]_{\bf{\bar{3}}} N &=& \frac{1}{\sqrt{2}} \m K_4^{(\varphi, \bar{\varphi}^{\prime},\bar{\varphi}^{})}
\nonumber \\
\left[ (N \bar{\varphi})_{\bf{\bar{3}}} (\varphi \bar{\varphi}^\prime)_{\bf{\bar{3}}} \right]_{\bf{\bar{3}}} N &=& \frac{1}{\sqrt{2}} \m K_5^{(\varphi, \bar{\varphi}^{},\bar{\varphi}^{\prime})} \nonumber \\
%%---------------------------------
%%%%%%%%%%%%%% (N \varphi) (\bar{\varphi}\bar{\varphi}) N
%%----------------------------------------
\left[ (N \varphi)_{\bf{3}} (\bar{\varphi} \bar{\varphi}^\prime)_{\bf{3_+}} \right]_{\bf{\bar{3}_\pm}} N &=& \frac{1}{2} \left( \pm \m K_2^{(\varphi, \bar{\varphi}^{},\bar{\varphi}^{\prime})} + \frac{1}{\sqrt{2}} \m K_4^{(\varphi, \bar{\varphi}^{},\bar{\varphi}^{\prime})} \right) +  (\bar{\varphi} \leftrightarrow \bar{\varphi}^\prime)
\nonumber \\
\left[ (N \varphi)_{\bf{3}} (\bar{\varphi} \bar{\varphi}^\prime)_{\bf{3_-}} \right]_{\bf{\bar{3}_\pm}} N &=& \frac{1}{2} \left( \pm \m K_2^{(\varphi, \bar{\varphi}^{},\bar{\varphi}^{\prime})} + \frac{1}{\sqrt{2}} \m K_4^{(\varphi, \bar{\varphi}^{},\bar{\varphi}^{\prime})} \right) -  (\bar{\varphi} \leftrightarrow \bar{\varphi}^\prime) 
\nonumber \\
\left[ (N \varphi)_{\bf{3}} (\bar{\varphi} \bar{\varphi}^\prime)_{\bf{\bar{3}}} \right]_{\bf{\bar{3}}} N &=& \frac{1}{\sqrt{2}} K_5^{(\varphi, \bar{\varphi}^{},\bar{\varphi}^{\prime})} 
\nonumber \\
\left[ (N \varphi)_{\bf{\bar{3}_\pm}} (\bar{\varphi} \bar{\varphi}^\prime)_{\bf{3_+}} \right]_{\bf{\bar{3}}} N &=& \frac{1}{2} \left( \m K_1^{(\varphi, \bar{\varphi}^{\prime},\bar{\varphi}^{})} + \frac{1}{\sqrt{2}}  \m K_4^{(\varphi, \bar{\varphi}^{},\bar{\varphi}^{\prime})} \right)  \pm (\bar{\varphi} \leftrightarrow \bar{\varphi}^\prime) 
\nonumber \\
\left[ (N \varphi)_{\bf{\bar{3}_\pm}} (\bar{\varphi} \bar{\varphi}^\prime)_{\bf{3_-}} \right]_{\bf{\bar{3}}} N &=& \frac{1}{2} \left( \m K_1^{(\varphi, \bar{\varphi}^{\prime},\bar{\varphi}^{})} - \frac{1}{\sqrt{2}}  \m K_4^{(\varphi, \bar{\varphi}^{},\bar{\varphi}^{\prime})} \right)  \pm (\bar{\varphi} \leftrightarrow \bar{\varphi}^\prime) 
\nonumber \\
\left[ (N \varphi)_{\bf{\bar{3}_\pm}} (\bar{\varphi} \bar{\varphi}^\prime)_{\bf{\bar{3}}} \right]_{\bf{\bar{3}}} N  &=& \pm \frac{1}{\sqrt{2}} \m K_3^{(\varphi, \bar{\varphi}^{},\bar{\varphi}^{\prime})} + \frac{1}{2} \m K_5^{(\varphi, \bar{\varphi}^{},\bar{\varphi}^{\prime})} \nonumber
\end{eqnarray}
%%%%%%%%%%%%%%%%%%%%%%%%%%%%%%%%%%%%%%%%%%%%%%%%%%%%%%%%%%%%%%%%%%%%%%%%%%%%%%%%%%%%%%%%%%

\newpage
\item There are three types of linearly independent $NN \bar{\varphi} \bar{\varphi}^\prime \bar{\varphi}^{\prime \prime}$ invariants: 

\begin{eqnarray}
\m L_1^{(\bar{\varphi},\bar{\varphi}^\prime,\bar{\varphi}^{\prime\prime})} &=& \left[ (N N)_{\bf{3}} (\bar{\varphi} \bar{\varphi}^\prime)_{\bf{\bar{3}}} \right]_{\bf{3}} \bar{\varphi}^{\prime \prime}  \longrightarrow  \frac{1}{\sqrt{3}} \left(
\begin{array}{ccc}
 \bar{\varphi }_3 \bar{\varphi }'_3 \bar{\varphi }''_1 & 0 & 0 \\
  & \bar{\varphi }_1 \bar{\varphi }'_1 \bar{\varphi }''_2 & 0 \\
  &  & \bar{\varphi }_2 \bar{\varphi }'_2 \bar{\varphi }''_3
\end{array} 
\right), \nonumber \\
\m L_2^{(\bar{\varphi},\bar{\varphi}^\prime,\bar{\varphi}^{\prime\prime})} &=& \left[ (N \bar{\varphi})_{\bf{3}} (\bar{\varphi}^\prime \bar{\varphi}^{\prime \prime})_{\bf{\bar{3}}} \right]_{\bf{\bar{3}}} N \longrightarrow \frac{1}{2\sqrt{3}} \left(
\begin{array}{ccc}
 0 & \bar{\varphi }_1 \bar{\varphi }'_1 \bar{\varphi }''_1 & \bar{\varphi }_3 \bar{\varphi }'_3 \bar{\varphi }''_3 \\
   & 0 & \bar{\varphi }_2 \bar{\varphi }'_2 \bar{\varphi }''_2 \\
   &   & 0
\end{array}
\right), \nonumber \\
\m L_3^{(\bar{\varphi},\bar{\varphi}^\prime,\bar{\varphi}^{\prime\prime})} &=& \left[ (N \bar{\varphi})_{\bf{\bar{3}}} (\bar{\varphi}^\prime \bar{\varphi}^{\prime \prime})_{\bf{\bar{3}}} \right]_{\bf{\bar{3}}} N \longrightarrow \frac{1}{2\sqrt{3}} \left(
\begin{array}{ccc}
 0 & \bar{\varphi }_2 \bar{\varphi }'_3 \bar{\varphi }''_3 & \bar{\varphi }_1 \bar{\varphi }'_2 \bar{\varphi }''_2 \\
   & 0 & \bar{\varphi }_3 \bar{\varphi }'_1 \bar{\varphi }''_1 \\
   &   & 0
\end{array}
\right). \nonumber 
\end{eqnarray}

\begin{eqnarray} \nonumber
%%-------------------------------------
%%%%%%%  (NN) (\bar{\varphi}\bar{\varphi}) \bar{\varphi}
%%--------------------------------
\left[ (N N)_{\bf{3}} (\bar{\varphi} \bar{\varphi}^\prime)_{\bf{3_\pm}} \right]_{\bf{3}} \bar{\varphi}^{\prime \prime}  &=& \frac{1}{\sqrt{2}} \left( \m L_1^{(\bar{\varphi}^{\prime\prime},\bar{\varphi}^{\prime}, \bar{\varphi}^{})} \pm \m L_1^{(\bar{\varphi}^{},\bar{\varphi}^{\prime\prime}, \bar{\varphi}^{\prime})} \right) 
\\ \nonumber
\left[ (N N)_{\bf{\bar{3}}} (\bar{\varphi} \bar{\varphi}^\prime)_{\bf{3_\pm}} \right]_{\bf{3}} \bar{\varphi}^{\prime \prime}  & =& \m L_3^{(\bar{\varphi}^{\prime},\bar{\varphi}^{}, \bar{\varphi}^{\prime\prime})} \pm (\bar{\varphi} \leftrightarrow \bar{\varphi}^\prime)
\\ \nonumber
\left[ (N N)_{\bf{\bar{3}}} (\bar{\varphi} \bar{\varphi}^\prime)_{\bf{\bar{3}}} \right]_{\bf{3_\pm}} \bar{\varphi}^{\prime \prime}  &=& \m L_2^{(\bar{\varphi}^{},\bar{\varphi}^{\prime}, \bar{\varphi}^{\prime\prime})} \pm \m L_3^{(\bar{\varphi}^{\prime\prime},\bar{\varphi}^{\prime}, \bar{\varphi}^{})}\\ \nonumber
%%----------------------------------------------------
%%%%%%%%%%%  (N \bar{\varphi}) (\bar{\varphi} \bar{\varphi}) N
%%-------------------------------------------------------------
\left[ (N \bar{\varphi}^{})_{\bf{3}} (\bar{\varphi}^\prime \bar{\varphi}^{\prime\prime})_{\bf{3_+}} \right]_{\bf{\bar{3}_\pm}} N &=& \frac{1}{2} \left( \m L_1^{(\bar{\varphi}^{},\bar{\varphi}^{\prime}, \bar{\varphi}^{\prime\prime})} + \m L_1^{(\bar{\varphi}^{},\bar{\varphi}^{\prime\prime}, \bar{\varphi}^{\prime})} \right) \pm \frac{1}{4} \left( \m L_3^{(\bar{\varphi}^{\prime\prime},\bar{\varphi}^{\prime}, \bar{\varphi}^{})} + \m L_3^{(\bar{\varphi}^{\prime},\bar{\varphi}^{}, \bar{\varphi}^{\prime\prime})} \right) \\ \nonumber
\left[ (N \bar{\varphi}^{})_{\bf{3}} (\bar{\varphi}^\prime \bar{\varphi}^{\prime\prime})_{\bf{3_-}} \right]_{\bf{\bar{3}_\pm}} N &=& \frac{1}{2} \left( - \m L_1^{(\bar{\varphi}^{},\bar{\varphi}^{\prime}, \bar{\varphi}^{\prime\prime})} + \m L_1^{(\bar{\varphi}^{},\bar{\varphi}^{\prime\prime}, \bar{\varphi}^{\prime})} \right) \pm \frac{1}{4} \left( \m L_3^{(\bar{\varphi}^{\prime\prime},\bar{\varphi}^{\prime}, \bar{\varphi}^{})} - \m L_3^{(\bar{\varphi}^{\prime},\bar{\varphi}^{}, \bar{\varphi}^{\prime\prime})} \right) \\ \nonumber
\left[ (N \bar{\varphi}^{})_{\bf{\bar{3}}} (\bar{\varphi}^\prime \bar{\varphi}^{\prime\prime})_{\bf{3_\pm}} \right]_{\bf{\bar{3}}} N &=& \frac{1}{\sqrt{2}} \left( \m L_3^{(\bar{\varphi}^{\prime\prime},\bar{\varphi}^{\prime}, \bar{\varphi}^{})} \pm \m L_3^{(\bar{\varphi}^{\prime},\bar{\varphi}^{}, \bar{\varphi}^{\prime\prime})} \right) \\ \nonumber
%%--------------------------------------
%%%%%%%%%%%  (N \bar{\varphi}) (N\bar{\varphi}) \bar{\varphi}
\left[ (N \bar{\varphi})_{\bf{3}} (N \bar{\varphi}^{\prime})_{\bf{3}} \right]_{\bf{3}}  \bar{\varphi}^{\prime \prime} &=& \m L_1^{(\bar{\varphi}^{},\bar{\varphi}^{\prime}, \bar{\varphi}^{\prime\prime})} \\ \nonumber
\left[ (N \bar{\varphi})_{\bf{3}} (N \bar{\varphi}^{\prime})_{\bf{\bar{3}}} \right]_{\bf{3}}  \bar{\varphi}^{\prime \prime} &=& \m L_3^{(\bar{\varphi}^{\prime\prime},\bar{\varphi}^{\prime}, \bar{\varphi}^{})} \\ \nonumber
\left[ (N \bar{\varphi})_{\bf{\bar{3}}} (N \bar{\varphi}^{\prime})_{\bf{\bar{3}}} \right]_{\bf{3_\pm}}  \bar{\varphi}^{\prime \prime} &=& \frac{1}{\sqrt{2}}  \m L_3^{(\bar{\varphi}^{\prime},\bar{\varphi}^{}, \bar{\varphi}^{\prime\prime})} \pm (\bar{\varphi} \leftrightarrow \bar{\varphi}^\prime)
\end{eqnarray}
\vskip .5cm
\item Invariants that contain intermediate $\bf{1}$, $\bf{1^\prime}$ and $\bf{\bar{1}^\prime}$ representations are given by:

\begin{eqnarray}
\left [ (NN)_{\bf{\bar{3}}}~ (\bar{\varphi}\varphi)_{\bf{1^\prime}} \right ]_{\bf{\bar{3}}} \varphi^\prime &\longrightarrow & \frac{(\bar{\varphi}\varphi)_{\bf{1^\prime}} }{\sqrt{6}} 
\begin{pmatrix}{
0 & \omega^2 \varphi_3^\prime & \omega \varphi_2^\prime \cr
& 0 & \varphi_1^\prime \cr
& & 0 
}\end{pmatrix} \nonumber \\
\left [ (NN)_{\bf{\bar{3}}}~ (\bar{\varphi}\varphi)_{\bf{\bar{1}^\prime}} \right ]_{\bf{\bar{3}}} \varphi^\prime &\longrightarrow & \frac{(\bar{\varphi}\varphi)_{\bf{\bar{1}^\prime}} }{\sqrt{6}} 
\begin{pmatrix}{
0 & \omega \varphi_3^\prime & \omega^2 \varphi_2^\prime \cr
& 0 & \varphi_1^\prime \cr
& & 0 
}\end{pmatrix} \nonumber  \\
\left [ (N\bar{\varphi})_{\bf{1}}~ (N\varphi)_{\bf{\bar{3}_\pm}} \right ]_{\bf{\bar{3}}} \varphi^\prime &=& \left [ (N\bar{\varphi})_{\bf{1}}~ (\varphi\varphi^\prime)_{\bf{\bar{3}_\pm}} \right ]_{\bf{\bar{3}}} N \nonumber \\
&=& \frac{1}{\sqrt{6}} \left( \m J_1^{(\varphi, \bar{\varphi}, \varphi^\prime)} + \frac{1}{\sqrt{2}} \m J_4^{(\varphi^\prime, \bar{\varphi}, \varphi)} + \frac{1}{\sqrt{2}} \m J_3^{(\varphi, \bar{\varphi}, \varphi^\prime)} \right) \pm (\varphi \leftrightarrow \varphi^\prime) 
\nonumber \\
\left [ (N\bar{\varphi})_{\bf{1^\prime}}~ (N\varphi)_{\bf{\bar{3}_\pm}} \right ]_{\bf{\bar{3}}} \varphi^\prime &=& \omega \left [ (N\bar{\varphi})_{\bf{1^\prime}}~ (\varphi\varphi^\prime)_{\bf{\bar{3}_\pm}} \right ]_{\bf{\bar{3}}} N \nonumber \\
&=& \frac{1}{\sqrt{6}} \left( \omega \m J_1^{(\varphi, \bar{\varphi}, \varphi^\prime)} + \frac{\omega^2}{\sqrt{2}}  \m J_4^{(\varphi^\prime, \bar{\varphi}, \varphi)} + \frac{1}{\sqrt{2}} \m J_3^{(\varphi, \bar{\varphi}, \varphi^\prime)} \right) \pm \omega (\varphi \leftrightarrow \varphi^\prime) 
\nonumber \\
\left [ (N\bar{\varphi})_{\bf{\bar{1}^\prime}}~ (N\varphi)_{\bf{\bar{3}_\pm}} \right ]_{\bf{\bar{3}}} \varphi^\prime &=& \omega^2 \left [ (N\bar{\varphi})_{\bf{\bar{1}^\prime}}~ (\varphi\varphi^\prime)_{\bf{\bar{3}_\pm}} \right ]_{\bf{\bar{3}}} N \nonumber \\
&=& \frac{1}{\sqrt{6}} \left( \omega^2 \m J_1^{(\varphi, \bar{\varphi}, \varphi^\prime)} + \frac{\omega}{\sqrt{2}}  \m J_4^{(\varphi^\prime, \bar{\varphi}, \varphi)} + \frac{1}{\sqrt{2}} \m J_3^{(\varphi, \bar{\varphi}, \varphi^\prime)} \right) \pm \omega^2 (\varphi \leftrightarrow \varphi^\prime) 
\nonumber \\
\left [ (N\varphi)_{\bf{\bar{3}_\pm}}~ (\bar{\varphi}\varphi^\prime)_{\bf{1^\prime}} \right ]_{\bf{\bar{3}}} N &\longrightarrow &  \frac{(\bar{\varphi}\varphi^\prime)_{\bf{1}^\prime} (\omega \pm 1)}{2\sqrt{6}} \begin{pmatrix}{
0 & \varphi_3 & \omega^2\varphi_2 \cr
& 0 & \omega \varphi_1 \cr
& & 0
}\end{pmatrix} \nonumber \\
\left [ (N\varphi)_{\bf{\bar{3}_\pm}}~ (\bar{\varphi}\varphi^\prime)_{\bf{\bar{1}^\prime}} \right ]_{\bf{\bar{3}}} N &\longrightarrow &  \frac{(\bar{\varphi}\varphi^\prime)_{\bf{\bar{1}}^\prime}(\omega^2\pm 1)}{2\sqrt{6}} \begin{pmatrix}{
0 & \varphi_3 & \omega\varphi_2 \cr
& 0 & \omega^2 \varphi_1 \cr
& & 0
}\end{pmatrix} \nonumber \\
\left[ (NN)_{\bf{3}} (\bar{\varphi} \varphi)_{\bf{1^\prime}} \right]_{\bf{3}} \bar{\varphi^\prime} &\longrightarrow &  \frac{(\bar{\varphi}\varphi)_{\bf{1}^\prime}}{\sqrt{3}} \begin{pmatrix}{
\omega \bar{\varphi}_2^\prime & 0 & 0 \cr
& \omega^2 \bar{\varphi}_3^\prime & 0 \cr
& & \bar{\varphi}_1^\prime
}\end{pmatrix} \nonumber \\
\left[ (NN)_{\bf{3}} (\bar{\varphi} \varphi)_{\bf{\bar{1}^\prime}} \right]_{\bf{3}} \bar{\varphi^\prime} &\longrightarrow &  \frac{(\bar{\varphi}\varphi)_{\bf{\bar{1}}^\prime}}{\sqrt{3}} \begin{pmatrix}{
\omega^2 \bar{\varphi}_2^\prime & 0 & 0 \cr
& \omega \bar{\varphi}_3^\prime & 0 \cr
& & \bar{\varphi}_1^\prime
}\end{pmatrix} \nonumber \\
\left[ (N \bar{\varphi})_{\bf{1}} (N \bar{\varphi}^\prime)_{\bf{\bar{3}}} \right]_{\bf{\bar{3}}} \varphi &=& \left[ (N \bar{\varphi})_{\bf{1}} (N \varphi)_{\bf{3}} \right]_{\bf{3}} \bar{\varphi}^\prime ~=~ \left[ (N \bar{\varphi})_{\bf{1}} (\varphi \bar{\varphi}^\prime)_{\bf{\bar{3}}} \right]_{\bf{\bar{3}}} N \nonumber \\
&=& \frac{1}{\sqrt{3}} \m K_2^{(\varphi, \bar{\varphi}^\prime, \bar{\varphi})} + \frac{1}{\sqrt{6}} \left( \m K_5^{(\varphi, \bar{\varphi}, \bar{\varphi}^\prime)} + \m K_4^{(\varphi, \bar{\varphi}, \bar{\varphi}^\prime)} \right)  
\nonumber \\
\left[ (N \bar{\varphi})_{\bf{1^\prime}} (N \bar{\varphi}^\prime)_{\bf{\bar{3}}} \right]_{\bf{\bar{3}}} \varphi  &=& \left[ (N \bar{\varphi})_{\bf{1^\prime}} (N \varphi)_{\bf{3}} \right]_{\bf{3}} \bar{\varphi}^\prime ~=~ \left[ (N \bar{\varphi})_{\bf{1^\prime}} (\varphi \bar{\varphi}^\prime)_{\bf{\bar{3}}} \right]_{\bf{\bar{3}}} N \nonumber \\
&=& \frac{1}{\sqrt{3}} \m K_2^{(\varphi, \bar{\varphi}^\prime, \bar{\varphi})} + \frac{1}{\sqrt{6}} \left( \omega^2 \m K_5^{(\varphi, \bar{\varphi}, \bar{\varphi}^\prime)} + \omega \m K_4^{(\varphi, \bar{\varphi}, \bar{\varphi}^\prime)} \right)   
\nonumber \\
\left[ (N \bar{\varphi})_{\bf{\bar{1}^\prime}} (N \bar{\varphi}^\prime)_{\bf{\bar{3}}} \right]_{\bf{\bar{3}}} \varphi &=& \left[ (N \bar{\varphi})_{\bf{\bar{1}^\prime}} (N \varphi)_{\bf{3}} \right]_{\bf{3}} \bar{\varphi}^\prime ~=~ \left[ (N \bar{\varphi})_{\bf{\bar{1}^\prime}} (\varphi \bar{\varphi}^\prime)_{\bf{\bar{3}}} \right]_{\bf{\bar{3}}} N \nonumber \\
&=& \frac{1}{\sqrt{3}} \m K_2^{(\varphi, \bar{\varphi}^\prime, \bar{\varphi})} + \frac{1}{\sqrt{6}} \left( \omega \m K_5^{(\varphi, \bar{\varphi}, \bar{\varphi}^\prime)} + \omega^2 \m K_4^{(\varphi, \bar{\varphi}, \bar{\varphi}^\prime)} \right)   
\nonumber \\
\left[ (N \bar{\varphi})_{\bf{\bar{3}}} (\varphi \bar{\varphi}^\prime)_{\bf{1^\prime}} \right]_{\bf{\bar{3}}} N &\longrightarrow & \frac{(\varphi\bar{\varphi}^\prime)_{\bf{1}^\prime}}{\sqrt{3}} \begin{pmatrix}{
\bar{\varphi}_2 & 0 & 0\cr
& \omega \bar{\varphi}_3 & 0 \cr
& & \omega^2 \bar{\varphi}_1
}\end{pmatrix}
\nonumber \\
\left[ (N \bar{\varphi})_{\bf{\bar{3}}} (\varphi \bar{\varphi}^\prime)_{\bf{\bar{1}^\prime}} \right]_{\bf{\bar{3}}} N &\longrightarrow & \frac{(\varphi\bar{\varphi}^\prime)_{\bf{\bar{1}}^\prime}}{\sqrt{3}} \begin{pmatrix}{
\bar{\varphi}_2 & 0 & 0\cr
& \omega^2 \bar{\varphi}_3 & 0 \cr
& & \omega \bar{\varphi}_1
}\end{pmatrix} 
\nonumber \\
\left[ (N \bar{\varphi})_{\bf{1}} (\bar{\varphi}^\prime \bar{\varphi}^{\prime \prime})_{\bf{\bar{3}}} \right]_{\bf{\bar{3}}} N &=&\left[ (N \bar{\varphi})_{\bf{1}} (N \bar{\varphi}^{\prime })_{\bf{3}} \right]_{\bf{3}} \bar{\varphi}^{\prime \prime} \nonumber \\
&=& \frac{1}{\sqrt{3}} \left( \m L_1^{(\bar{\varphi}^{\prime\prime},\bar{\varphi}^\prime, \bar{\varphi})} + \m L_2^{(\bar{\varphi}^{},\bar{\varphi}^\prime, \bar{\varphi}^{\prime\prime})} + \m L_3^{(\bar{\varphi}^{},\bar{\varphi}^\prime, \bar{\varphi}^{\prime\prime})} \right) 
\nonumber \\
\left[ (N \bar{\varphi})_{\bf{1^\prime}} (\bar{\varphi}^\prime \bar{\varphi}^{\prime \prime})_{\bf{\bar{3}}} \right]_{\bf{\bar{3}}} N &=& \left[ (N \bar{\varphi})_{\bf{1^\prime}} (N \bar{\varphi}^{\prime })_{\bf{3}} \right]_{\bf{3}} \bar{\varphi}^{\prime \prime} \nonumber \\
&=& \frac{1}{\sqrt{3}} \left( \m L_1^{(\bar{\varphi}^{\prime\prime},\bar{\varphi}^\prime, \bar{\varphi})} + \omega \m L_2^{(\bar{\varphi}^{},\bar{\varphi}^\prime, \bar{\varphi}^{\prime\prime})} + \omega^2 \m L_3^{(\bar{\varphi}^{},\bar{\varphi}^\prime, \bar{\varphi}^{\prime\prime})} \right) 
\nonumber \\
\left[ (N \bar{\varphi})_{\bf{\bar{1}^\prime}} (\bar{\varphi}^\prime \bar{\varphi}^{\prime \prime})_{\bf{\bar{3}}} \right]_{\bf{\bar{3}}} N &=&\left[ (N \bar{\varphi})_{\bf{\bar{1}^\prime}} (N \bar{\varphi}^{\prime })_{\bf{3}} \right]_{\bf{3}} \bar{\varphi}^{\prime \prime} \nonumber \\
&=& \frac{1}{\sqrt{3}} \left( \m L_1^{(\bar{\varphi}^{\prime\prime},\bar{\varphi}^\prime, \bar{\varphi})} +\omega^2 \m L_2^{(\bar{\varphi}^{},\bar{\varphi}^\prime, \bar{\varphi}^{\prime\prime})} +\omega \m L_3^{(\bar{\varphi}^{},\bar{\varphi}^\prime, \bar{\varphi}^{\prime\prime})} \right) 
\nonumber 
\end{eqnarray}
\end{itemize}

%%%%%%%%%%%%%%%%%

\vskip .5cm

\section*{Appendix B: The Special Matrix and Dimension-Six Invariants}

As mentioned in the main text, aside from the singular coupling, there exist other dimension-six invariants capable of producing the special matrix. These fall into two distinct classes, which we detail below. 

\begin{enumerate}[(I)]

\item The first class is characterized by the fact that it gives constraints which are independent of the vacuum values of the familon fields, giving a prediction for the value of $r$. For example, we find the monomial arrangement

$$\begin{pmatrix}{ 2\varphi^{}_3\varphi^{\prime}_3\varphi^{\prime\prime}_3&\pm\varphi^{}_2\varphi^{\prime}_1\varphi^{\prime\prime}_1&\pm\varphi^{}_1\varphi^{\prime}_3\varphi^{\prime\prime}_3 \cr
& 2\varphi^{}_1\varphi^{\prime}_1\varphi^{\prime\prime}_1&\pm\varphi^{}_3\varphi^{\prime}_2\varphi^{\prime\prime}_2 \cr
& & 2\varphi^{}_2\varphi^{\prime}_2\varphi^{\prime\prime}_2
}\end{pmatrix},
$$
produced by the invariant $\left[ (N\varphi)_{\bf{\bar{3}_\pm}}(\varphi^\prime\varphi^{\prime\prime})_{\bf{3}} \right]_{\bf{\bar{3}}} N$.  It is capable of producing the special matrix, but subject to the constraint 

$$a^{}_{11}a^{}_{22}a^{}_{33}~=~\pm 8a^{}_{12}a^{}_{13}a^{}_{23}.$$
Unfortunately, with the additional TBM constraints, all such couplings yield ($r=\pm 1/8$), which is incompatible with the data. 

\item This class is characterized by their complicated structure, and leads to under-constrained systems. An example is $\left[ (N\varphi)_{\bf{\bar{3}}_+} (N\varphi^\prime)_{\bf{3}_+} \right]_{\bf{3}_+} \bar{\varphi}$, which gives the matrix

$$\begin{pmatrix}{
2 \bar{\varphi}_1 A_{23} & \bar{\varphi}_1 A_{13} + \bar{\varphi}_2 A_{23} + 2 \varphi_3 \varphi_3^\prime \bar{\varphi}_3 & \bar{\varphi}_1 A_{12} + \bar{\varphi}_3 A_{23} + 2 \varphi_2 \varphi_2^\prime \bar{\varphi}_2 \cr & 2 \bar{\varphi}_2 A_{13} & \bar{\varphi}_2 A_{12} + \bar{\varphi}_3 A_{13} + 2 \varphi_1 \varphi_1^\prime \bar{\varphi}_1 \cr & & 2 \bar{\varphi}_3 A_{12}
}\end{pmatrix}, 
$$
written in terms of 
$$
A_{ij} = \varphi_i \varphi^\prime_j + \varphi_i^\prime \varphi_j. $$ 

We find two types of particular solutions that can reproduce the special matrix. The first one is obtained by setting $\langle \varphi_1 \rangle =0$, and the familon vacuum structure is then given by

\begin{eqnarray}
\begin{pmatrix}{\varphi_1^{}\cr\varphi_2^{}\cr\varphi_3^{}}\end{pmatrix} &=& \begin{pmatrix}{0\cr1\cr-\lambda^{4}}\end{pmatrix}\varphi^{}_2,\nonumber \\
\begin{pmatrix}{\varphi_1'\cr\varphi_2'\cr\varphi_3'}\end{pmatrix} &=& \begin{pmatrix}{1\cr-\frac{\sqrt{2r}}{4}(1+2\sqrt{2r})\lambda^4\cr \frac{\sqrt{2r}}{4}(1+2\sqrt{2r})\lambda^8}\end{pmatrix}{\varphi_1'},\nonumber \\
 \begin{pmatrix}{\overline \varphi_1^{}\cr\overline \varphi_2^{}\cr\overline \varphi_3^{}}\end{pmatrix} &=& \begin{pmatrix}{-\sqrt{2r}{\lambda^8}\cr-\lambda^4\cr1}\end{pmatrix}\overline \varphi^{}_3. \nonumber
\end{eqnarray}

The second set of solutions have no familon components of zero vacuum value, which mimics the solution for the singular invariant in the main text. The familon vacuum structure of this solution is given by

\begin{eqnarray}
\begin{pmatrix}{\varphi_1^{}\cr\varphi_2^{}\cr\varphi_3^{}}\end{pmatrix} &=& \begin{pmatrix}{1 \cr \alpha \lambda^4 \cr \alpha \lambda^8}\end{pmatrix}\varphi^{}_1,\nonumber \\
\begin{pmatrix}{\varphi_1'\cr\varphi_2'\cr\varphi_3'}\end{pmatrix} &=& \begin{pmatrix}{1 \cr \alpha^\prime \lambda^4 \cr \alpha^\prime \lambda^8}\end{pmatrix}{\varphi_1'},\nonumber \\
 \begin{pmatrix}{\overline \varphi_1^{}\cr\overline \varphi_2^{}\cr\overline \varphi_3^{}}\end{pmatrix} &=& \begin{pmatrix}{\bar{\alpha} \lambda^8 \cr \lambda^4 \cr 1}\end{pmatrix}\overline \varphi^{}_3, \nonumber
\end{eqnarray}
subject to the conditions
\begin{eqnarray}
\frac{1}{\alpha} + \frac{1}{\alpha^\prime} = 4 \left(1-\frac{1}{\bar{\alpha}} \right), \quad\quad r= \frac{\bar{\alpha}^2}{2(\bar{\alpha}-1)}. \nonumber
\end{eqnarray}

\end{enumerate}

%%%%%%%%%%%%%
\newpage
\section*{Appendix C: An Exemplar $Y^{(-1)}$}

A complete description of the neutrino mixing angles in $\m U_{MNSP}$,

\be \nonumber
\m U^{}_{MNSP}~=~\m U^\dagger_{-1}\,\m U^{}_{\rm seesaw},\ee 
requires knowledge of the matrix which diagonalizes $Y^{(-1)}$. Depending on the particular seesaw mixing scheme, $\m U_{\rm seesaw}$, these corrections can be as large as $\m O (\lambda)$.  
 
A particularly intriguing possibility for generating such corrections arises from extending the program of simultaneously considering family symmetries and ideas from grand unification.  In the main text we have explored taking $Y^{(0)}$ proportional to $Y^{(2/3)}$ and the consequences this form of $Y^{(0)}$ has for the neutrino Majorana matrix.  However, GUTs based on $SU(5)$ also predict that, up to the insertion of Georgi-Jarlskog factors\cite{Georgi:1979df}, the charged-lepton and down-quark Dirac matrices are related by a transpose, $Y^{(-1)}\sim Y^{(-1/3)T}$.

As a specific example of how such a scheme may work in principle, suppose, as in our model, that $Y^{(2/3)}$ is diagonal.  In this case, 

\be\label{ckm}
\m U_d ~=~ \m U^{}_{CKM} ~=~\begin{pmatrix}{
1-\lambda^2/2 & \lambda & A \lambda^3 (\rho - i \eta) \cr -\lambda & 1 - \lambda^2/2 & A \lambda^2 \cr  A \lambda^3 (1 - \rho - i \eta) & -A \lambda^2 & 1 
}\end{pmatrix}.
\ee
This implies that the down-quark Dirac matrix $Y^{(-1/3)}$ is known up to a unitary right-handed rotation matrix, $\m V$, 

\be
 Y^{(-1/3)}_{} = \m U_{CKM}^{} \m D^{}_d\, \m V_{}^{\dagger}.
\ee
Surprisingly, one may check that for 

\be 
\m D_d ~=~ m_b \begin{pmatrix}{
- \frac{\lambda^4}{3} & & \cr & + \frac{\lambda^2}{3} & \cr & & 1
}\end{pmatrix}, 
\ee
with GUT scale values\cite{ross&serna}, $\lambda = 0.227$,  $\rho = 0.22$, and $\eta = 0.33$, and

\be 
\m V ~=~ \begin{pmatrix}{
 \cos \beta_{13} & 0 & \sin \beta_{13}\cr 0 & 1 & 0 \cr -\sin \beta_{13} & 0 & \cos \beta_{13}
}\end{pmatrix}, \qquad \beta_{13} = 3^\circ \approx \lambda^2 ,
\ee
a suitable $Y^{(-1)}$ can be found. For this form of $\m V$, one finds that $Y^{(-1/3)}$ is given by,

\be
 Y^{(2/3)}_{} \sim \begin{pmatrix}{
-\frac{1}{3} \lambda^4 + 
 A \lambda^5 (\rho - i \eta ) & \lambda^3 / 3 &  A \lambda^3 (\rho - i \eta )  \cr A \lambda^4 + \lambda^5/3 & \frac{1}{3} \lambda^2 (1 - \lambda^2/2) & A \lambda^2 \cr \lambda^2 & - A \lambda^4 /3 & 1}
\end{pmatrix} + \m O(\lambda^6).
\ee 
Assuming the $SU(5)$ relation $Y^{(-1)}\sim Y^{(-1/3)T}$ holds, and that the $(22)$ and $(23)$ elements of $Y^{(-1/3)}$ are generated by a $\bf{45}$ Higgs which gives additional Georgi-Jarlskog factors of $-3$ to the $(22)$ and $(32)$ elements of $Y^{(-1)}$, we have,

\be
 Y^{(-1)}_{} \sim \begin{pmatrix}{
-\frac{1}{3} \lambda^4 + 
 A \lambda^5 (\rho - i \eta ) & A \lambda^4 + \lambda^5 /3 &  \lambda^2  \cr \lambda^3 /3 & - \lambda^2 (1 - \lambda^2/2) & -A \lambda^4/3 \cr  A \lambda^3 (\rho - i \eta ) &  -3A \lambda^2  & 1}
\end{pmatrix} + \m O(\lambda^6).
\ee
For $0.72 < A < 0.74$, and assuming TBM mixing in $\m U_{\rm seesaw}$, one finds that diagonalization of $Y^{(-1)}$ yields, 

\begin{eqnarray}
 && 30.9^\circ <\theta_{12}< 31^\circ , \qquad \theta_{23} = 44.7^\circ , \qquad  8.34^\circ< \theta_{13} < 8.50^\circ , \cr 
 && 0.00462<\frac{m_e}{m_{\mu}}<0.00495 , \qquad \qquad \frac{m_\mu}{m_{\tau}}=0.0504,
\end{eqnarray}
giving charged-lepton mass ratios in close agreement with their GUT scale values \cite{ross&serna}, and reasonable neutrino mixing angles with respect to their global fits \cite{fit}.

At this stage this is a numerical proof-of-principle. In a future work \cite{nextpaper}, we hope to realize such matrices using $\bs{\m Z_7 \rtimes \m Z_3}$.

\newpage  
%%%%%%%%%%%%%
{}

%%%%%%%%%%%%%%%
\end{document}